\documentclass{article}

\usepackage{arxiv}

\usepackage[utf8]{inputenc} 
\usepackage[T1]{fontenc}    
\usepackage{lmodern}        
\usepackage{hyperref}       
\usepackage{url}            
\usepackage{booktabs}       
\usepackage{amsfonts}       
\usepackage{nicefrac}       
\usepackage{microtype}      
\usepackage{graphicx}

\title{Simulating High-Dimensional Multivariate Data using the Bigsimr R
Package}

\author{
    A. Grant Schissler
   \\
    Department of Mathematics \& Statistics \\
    University of Nevada, Reno \\
   \\
  \texttt{} \\
   \And
    Edward J. Bedrick
   \\
    Epidemiology and Biostatistics Department \\
    University of Arizona \\
   \\
  \texttt{} \\
   \And
    Alexander D. Knudson
   \\
    Department of Mathematics \& Statistics \\
    University of Nevada, Reno \\
   \\
  \texttt{} \\
   \And
    Tomasz J. Kozubowski
   \\
    Department of Mathematics \& Statistics \\
    University of Nevada, Reno \\
   \\
  \texttt{} \\
   \And
    Tin Nguyen
   \\
    Department of Computer Science \& Engineering \\
    University of Nevada, Reno \\
   \\
  \texttt{} \\
   \And
    Anna K. Panorska
   \\
    Department of Mathematics \& Statistics \\
    University of Nevada, Reno \\
   \\
  \texttt{} \\
   \And
    Juli Petereit
   \\
    Nevada Bioinformatics Center \\
    University of Nevada, Reno \\
   \\
  \texttt{} \\
   \And
    Walter W. Piegorsch
   \\
    Mathematics Department \\
    University of Arizona \\
   \\
  \texttt{} \\
   \And
    Duc Tran
   \\
    Department of Computer Science \& Engineering \\
    University of Nevada, Reno \\
   \\
  \texttt{} \\
  }

\usepackage{color}
\usepackage{fancyvrb}

\DefineVerbatimEnvironment{Highlighting}{Verbatim}{commandchars=\\\{\}}
\usepackage{framed}
\definecolor{shadecolor}{RGB}{248,248,248}
\newenvironment{Shaded}{\begin{snugshade}}{\end{snugshade}}

\newcommand{\AttributeTok}[1]{\textcolor[rgb]{0.77,0.63,0.00}{#1}}

\newcommand{\CommentTok}[1]{\textcolor[rgb]{0.56,0.35,0.01}{\textit{#1}}}

\newcommand{\ConstantTok}[1]{\textcolor[rgb]{0.00,0.00,0.00}{#1}}
\newcommand{\ControlFlowTok}[1]{\textcolor[rgb]{0.13,0.29,0.53}{\textbf{#1}}}

\newcommand{\DecValTok}[1]{\textcolor[rgb]{0.00,0.00,0.81}{#1}}
\newcommand{\DocumentationTok}[1]{\textcolor[rgb]{0.56,0.35,0.01}{\textbf{\textit{#1}}}}

\newcommand{\FloatTok}[1]{\textcolor[rgb]{0.00,0.00,0.81}{#1}}
\newcommand{\FunctionTok}[1]{\textcolor[rgb]{0.00,0.00,0.00}{#1}}

\newcommand{\NormalTok}[1]{#1}

\newcommand{\OtherTok}[1]{\textcolor[rgb]{0.56,0.35,0.01}{#1}}

\newcommand{\SpecialCharTok}[1]{\textcolor[rgb]{0.00,0.00,0.00}{#1}}

\newcommand{\StringTok}[1]{\textcolor[rgb]{0.31,0.60,0.02}{#1}}

\newlength{\csllabelwidth}
\newlength{\cslhangindent}
  {}%
  {\par}
\newenvironment{CSLReferences}[2] 
 {
  \setlength{\parindent}{0pt}
  \ifodd #1 \everypar{\setlength{\hangindent}{\cslhangindent}}\ignorespaces\fi
  \ifnum #2 > 0
  \setlength{\parskip}{#2\baselineskip}
  \fi
 }%
 {}
\usepackage{calc} 

\usepackage{amsmath}
\usepackage{amssymb}
\usepackage{amsfonts}
\usepackage{amsthm}

\usepackage{booktabs}
\usepackage{longtable}
\usepackage{array}
\usepackage{ragged2e}
\usepackage{setspace}

\begin{document}
\maketitle

\def\tightlist{}

© A.G. Schissler. This is an Open Access article distributed under the terms of the Creative Commons License Attribution-NonCommercial-ShareAlike 4.0 International (CC BY-NC-SA 4.0) (https://creativecommons.org/licenses/by-nc-sa/4.0/), which permits reusers to distribute, remix, adapt, and build upon the material in any medium or format for noncommercial purposes only, and only so long as attribution is given to the creator. If you remix, adapt, or build upon the material, you must license the modified material under identical terms.

\clearpage

\begin{abstract}
It is critical to accurately simulate data when employing Monte Carlo
techniques and evaluating statistical methodology. Measurements are
often correlated and high dimensional in this era of big data, such as
data obtained in high-throughput biomedical experiments. Due to the
computational complexity and a lack of user-friendly software available
to simulate these massive multivariate constructions, researchers resort
to simulation designs that posit independence or perform arbitrary data
transformations. To close this gap, we developed the Bigsimr Julia
package with R and Python interfaces. This paper focuses on the R
interface. These packages empower high-dimensional random vector
simulation with arbitrary marginal distributions and dependency via a
Pearson, Spearman, or Kendall correlation matrix. bigsimr contains
high-performance features, including multi-core and
graphical-processing-unit-accelerated algorithms to estimate correlation
and compute the nearest correlation matrix. Monte Carlo studies quantify
the accuracy and scalability of our approach, up to \(d=10,000\). We
describe example workflows and apply to a high-dimensional data set ---
RNA-sequencing data obtained from breast cancer tumor samples.
\end{abstract}


\hypertarget{introduction}{%
\section{Introduction}\label{introduction}}

Massive high-dimensional (HD) data sets are now common in many areas of
scientific inquiry. As new methods are developed for data analysis, a
fundamental challenge lies in designing and conducting simulation
studies to assess the operating characteristics of proposed methodology
--- such as false positive rates, statistical power, interval coverage,
and robustness. Further, efficient simulation empowers statistical
computing strategies, such as the parametric bootstrap (Chernick 2008)
to simulate from a hypothesized null model, providing inference in
analytically challenging settings. Such Monte Carlo (MC) techniques
become difficult for HD dependent data using existing algorithms and
tools. This is particularly true when simulating massive multivariate,
non-normal distributions, arising in many fields of study.

As others have noted, it can be vexing to simulate dependent,
non-normal/discrete data, even for low-dimensional (LD) settings (Madsen
and Birkes 2013; Xiao and Zhou 2019). For continuous non-normal LD
multivariate data, the well-known NORmal To Anything (NORTA) algorithm
(Cario and Nelson 1997) and other copula approaches (Nelsen 2007) are
well-studied and implemented in publicly-available software (Yan 2007;
Chen 2001). Yet these approaches do not scale in a timely fashion to HD
problems (X. Li et al. 2019). For discrete data, early simulation
strategies had major flaws, such as failing to obtain the full range of
possible correlations --- such as admitting only positive correlations
(Park, Park, and Shin 1996). While more recent approaches (Madsen and
Birkes 2013; Xiao 2017; Barbiero and Ferrari 2017) have remedied this
issue for LD problems, the existing tools are not designed to scale to
high dimensions.

Another central issue lies in characterizing dependence between
components in the HD random vector. The choice of correlation in
practice usually relates to the eventual analytic goal and
distributional assumptions of the data (e.g., non-normal, discrete,
infinite support, etc.). For normal data, the Pearson product-moment
correlation describes the dependence perfectly. However, simulating
arbitrary random vectors that match a target Pearson correlation matrix
is computationally intense (Chen 2001; Xiao 2017). On the other hand, an
analyst might consider use of nonparametric correlation measures to
better characterize monotone, non-linear dependence, such as Spearman's
\(\rho\) and Kendall's \(\tau\). Throughout, we focus on matching these
nonparametric dependence measures, as our aim lies in modeling
non-normal data and these rank-based measures possess invariance
properties enabling our proposed methodology. We do, however, implement
Pearson matching, but several layers of approximation are required.

With all this in mind, we present a scalable, flexible multivariate
simulation algorithm. The crux of the method lies in the construction of
a Gaussian copula in the spirit of the NORTA procedure. As we will
describe in more detail, the algorithm's design leverages useful
properties of nonparametric correlation measures, namely invariance
under monotone transformation and well-known closed-form relationships
between dependence measures for the multivariate normal (MVN)
distribution. For our method, we developed a high-performance
implementation: the \texttt{Bigsimr} Julia package, with R and Python
interfaces \texttt{bigsimr}.

This article proceeds by providing background information, including a
description of a motivating example application: RNA-sequencing
(RNA-seq) breast cancer data. Then we describe and justify our
simulation methodology and related algorithms. We proceed by providing
an illustrative LD \texttt{bigsimr} workflow. Next we conduct MC studies
under various bivariate distributional assumptions to evaluate
performance and accuracy. After the MC evaluations, we simulate random
vectors motivated by our RNA-seq example, evaluate the accuracy, and
provide example statistical computing tasks, namely MC estimation of
joint probabilities and evaluating HD correlation estimation efficiency.
Finally, we discuss the method's utility, limitations, and future
directions.

\hypertarget{background}{%
\section{Background}\label{background}}

The \texttt{Bigsimr} package presented here provides multiple
high-performance algorithms that operate with HD multivariate data. All
these algorithms were originally designed to support a single task: to
generate random vectors drawn from multivariate probability
distributions with given marginal distributions and dependency metrics.
Specifically, our goal is to efficiently simulate a large number, \(B\),
of HD random vectors \({\bf Y}=(Y_1, \ldots, Y_d)^\top\) with
\emph{correlated} components and heterogeneous marginal distributions,
described via cumulative distribution functions (CDFs)
\(F_i, i=1,\ldots,d\).

When designing this methodology, we developed the following properties
to guide our effort. We divide the properties into two categories: (1)
basic properties (BP) and `scalability' properties (SP). The BPs are
adapted from an existing criteria due to Nikoloulopoulos (2013). A
suitable simulation strategy should possess the following properties:

\begin{itemize}
\tightlist
\item
  BP1: A wide range of dependences, allowing both positive and negative
  values, and, ideally, admitting the full range of possible values.
\item
  BP2: Flexible dependence, meaning that the number of dependence
  parameters can be equal to the number of bivariate pairs in the
  vector.
\item
  BP3: Flexible marginal modeling, generating heterogeneous data ---
  including mixed continuous and discrete marginal distributions.
\end{itemize}

Moreover, the simulation method must \emph{scale} to high dimensions:

\begin{itemize}
\tightlist
\item
  SP1: Procedure must scale to high dimensions with practical
  computation times.
\item
  SP2: Procedure must scale to high dimensions while maintaining
  accuracy.
\end{itemize}

\hypertarget{motivating-example-rna-seq-data}{%
\subsection{Motivating example: RNA-seq
data}\label{motivating-example-rna-seq-data}}

Simulating HD, non-normal, correlated data motivates this work, in
pursuit of modeling RNA-sequencing (RNA-seq) data (Wang, Gerstein, and
Snyder 2009; Conesa et al. 2016) derived from breast cancer patients.
The RNA-seq data-generating process involves counting how often a
particular form of messenger RNA (mRNA) is expressed in a biological
sample. RNA-seq platforms typically quantify the entire transcriptome in
one experimental run, resulting in HD data. For human-derived samples,
this results in count data corresponding to over 20,000 genes
(protein-coding genomic regions) or even over 77,000 isoforms when
alternatively-spliced mRNA are counted (Schissler et al. 2019).
Importantly, due to inherent biological processes, gene expression data
exhibit correlation (co-expression) across genes (Efron 2007; Schissler,
Piegorsch, and Lussier 2018).

\begin{figure}
\centering
\includegraphics{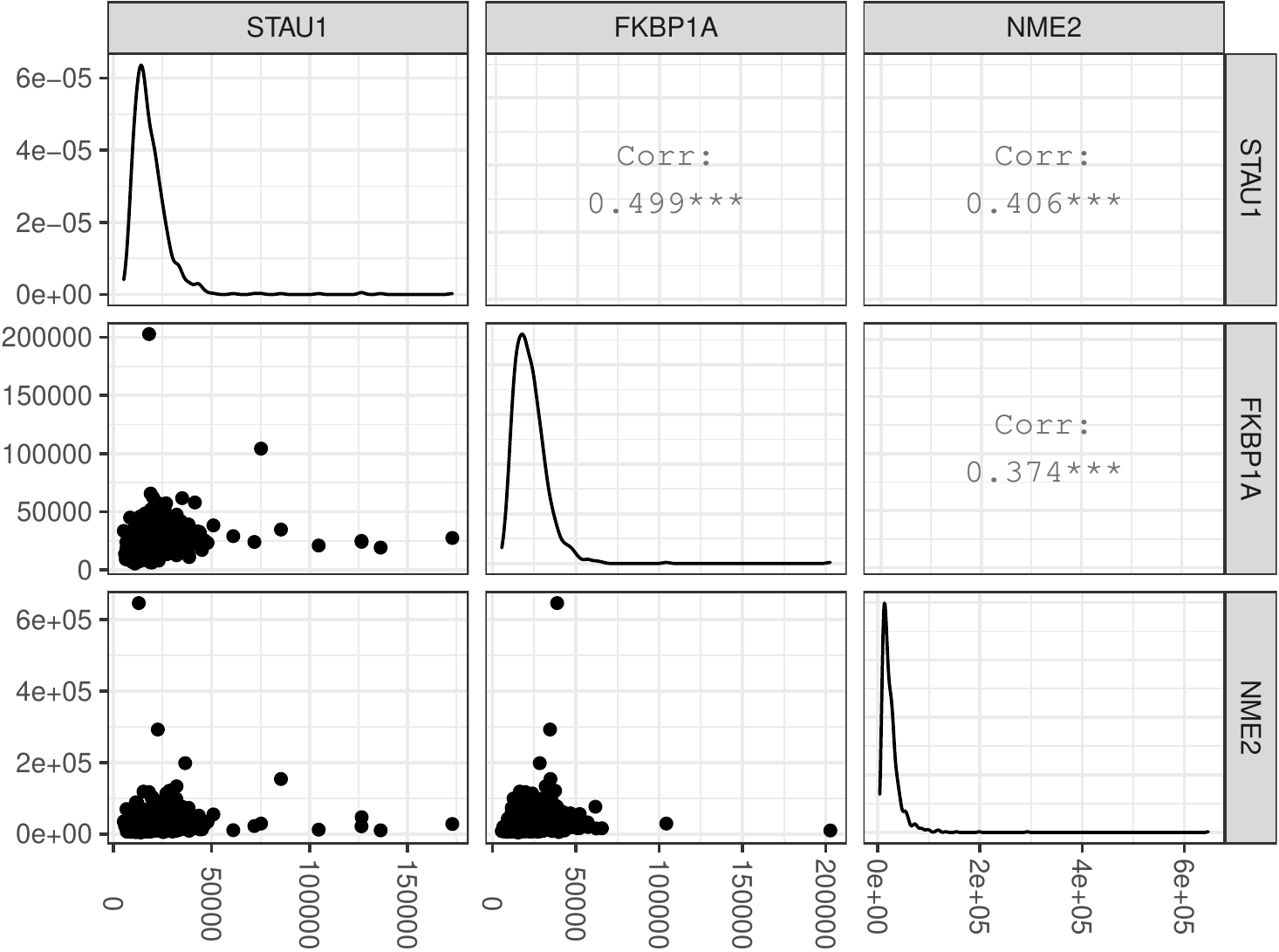}
\caption{\label{fig:ch010-realDataFig}Pairwise scatterplots for three
example genes (Table 1), with estimated Spearman's correlations and
marginal density estimates displayed. The data possess outliers,
heavy-right tails, are discrete, and have non-trivial intergene
correlations. Modeling these data motivate our simulation methodology.}
\end{figure}

We illustrate our methodology using the Breast Invasive Carcinoma (BRCA)
data set housed in The Cancer Genome Atlas (TCGA; see Acknowledgments).
For ease of modeling and simplicity of exposition, we only consider high
expressing genes. In turn, we begin by filtering to retain the top r d
of the highest-expressing genes (in terms of median expression) of the
over 20,000 gene measurements from \(N=r nrow(example_brca)\) patients'
tumor samples. This gives a great number of pairwise dependencies among
the marginals (specifically, \(r choose(d,2)\) correlation parameters).
Table \ref{tab:ch010-realDataTab} displays RNA-seq counts for three
selected high-expressing genes for the first five patients' breast tumor
samples. To help visualize the bivariate relationships for these three
selected genes across all patients, Figure \ref{fig:ch010-realDataFig}
displays the marginal distributions and estimated Spearman's
correlations.

\begin{table}

\caption{\label{tab:ch010-realDataTab}mRNA expression for three selected high-expressing genes, STAU1, FKBP1A, NME2, for the first five patients in the TCGA BRCA data set.}
\centering
\begin{tabular}[t]{lrrr}
\toprule
Patient ID & STAU1 & FKBP1A & NME2\\
\midrule
TCGA-A1-A0SB & 10440 & 11354 & 17655\\
TCGA-A1-A0SD & 21523 & 20221 & 14653\\
TCGA-A1-A0SE & 21733 & 22937 & 35251\\
TCGA-A1-A0SF & 11866 & 19650 & 16551\\
TCGA-A1-A0SG & 12486 & 12089 & 10434\\
\bottomrule
\end{tabular}
\end{table}

\hypertarget{measures-of-dependency}{%
\subsection{Measures of dependency}\label{measures-of-dependency}}

In multivariate analysis, an analyst must select a metric to quantify
dependency. The most widely-known is the Pearson correlation coefficient
that describes the linear association between two random variables \(X\)
and \(Y\), and, it is given by

\begin{equation}
\rho_P(X,Y) = \frac{E(XY) - E(X)E(Y)}{\left[ \mathrm{Var}(X)\mathrm{Var}(Y)\right]^{1/2}}.
\label{eq:pearson}
\end{equation}

As Madsen and Birkes (2013) and Mari and Kotz (2001) discuss, for a
bivariate normal \((X,Y)\) random vector, the Pearson correlation
completely describes the dependency between the components. For
non-normal marginals with monotone correlation patterns, however,
\(\rho_P\) suffers some drawbacks and may mislead or fail to capture
important relationships (Mari and Kotz 2001). Alternatively, analysts
often prefer rank-based correlation measures to describe the degree of
monotonic association.

Two nonparametric, rank-based measures common in practice are Spearman's
correlation (denoted \(\rho_S\)) and Kendall's \(\tau\). Spearman's
\(\rho_S\) has an appealing correspondence as the Pearson correlation
coefficient on \emph{ranks} of the values, thereby capturing nonlinear
yet monotone relationships. Kendall's \(\tau\), on the other hand, is
the difference in probabilities of concordant and discordant pairs of
observations, \((X_i, Y_i)\) and \((X_j, Y_j)\) (concordance meaning
that orderings have the same direction, e.g., if \(X_i < X_j\), then
\(Y_i < Y_j\)). Note that concordance is determined by the ranks of the
values, not the values themselves. Both \(\tau\) and \(\rho_S\) are
\emph{invariant under monotone transformations} of the underlying random
variates. As we will describe more fully in the
\protect\hyperlink{algorithms}{Algorithms} section, this property
enables matching rank-based correlations with speed (SP1) and accuracy
(SP2).

\emph{Correspondence among Pearson, Spearman, and Kendall correlations}

There is no closed form, general correspondence among the rank-based
measures and the Pearson correlation coefficient, as the marginal
distributions \(F_i\) are intrinsic in their calculation. For
\emph{bivariate normal vectors}, however, the correspondence is
well-known:

\begin{equation}
\label{eq:convertKendall}
\rho_{P} = \sin \left( \tau \times \frac{\pi}{2} \right), 
\end{equation}

\noindent and similarly for Spearman's \(\rho\) (Kruskal 1958),

\begin{equation}
\label{eq:convertSpearman}
\rho_P = 2 \times \sin \left( \rho_S \times \frac{\pi}{6} \right).
\end{equation}

\emph{Marginal-dependent bivariate correlation bounds}

Given two marginal distributions, \(\rho_P\) is not free to vary over
the entire range of possible correlations \([-1,1]\). The
\emph{Frechet-Hoeffding bounds} are well-studied (Nelsen 2007; Barbiero
and Ferrari 2017). These constraints cannot be overcome through
algorithm design. In general, the bounds are given by

\begin{equation}
\label{eq:frechet}
\rho_P^{max} = \rho_P \left( F^{-1}_1 (U), F^{-1}_2 (U) \right), \quad \rho_P^{min} = \rho_P \left( F^{-1}_1 (U), F^{-1}_2 (1 - U) \right)
\end{equation}

\noindent where \(U\) is a uniform random variable on \((0,1)\) and
\(F^{-1}_1, F^{-1}_2\) are the inverse CDFs of \(X_1\) and \(X_2\),
respectively, defined as

\begin{equation}
F_{i}^{-1} = \inf\{y:F_{i}(y) \geq u \}
\label{eq:inverseCDF}
\end{equation}

when the variables are discrete.

\hypertarget{gaussian-copulas}{%
\subsection{Gaussian copulas}\label{gaussian-copulas}}

There is a connection of our simulation strategy to Gaussian
\emph{copulas} (Nelsen 2007). A copula is a distribution function on
\([0,1]^d\) that describes a multivariate probability distribution with
standard uniform marginals. This provides a powerful, natural way to
characterize joint probability. Consequently, the study of copulas is an
important and active area of statistical theory and practice. For any
random vector \({\bf X}=(X_1, \ldots, X_d)^\top\) with CDF \(F\) and
marginal CDFs \(F_i\) there is a copula function \(C(u_1, \ldots, u_d)\)
satisfying

\begin{equation}
F(x_1, \ldots, x_d) = {\mathbb P}(X_1\leq x_1, \ldots,X_d\leq x_d) = C(F_1(x_1), \ldots, F_d(x_d)), \quad x_i \in {\mathbb R}, i=1,\ldots,d.
\label{eq:copula}
\end{equation}

A Gaussian copula has marginal CDFs that are all standard normal,
\(F_i = \Phi, \forall \, i\), . This representation corresponds to a
multivariate normal (MVN) distribution with standard normal marginal
distributions and covariance matrix \(R_P\). Since the marginals are
standardized to have unit variance, this \(R_P\) is also Pearson
correlation matrix. If \(F_R\) is the CDF of such a multivariate normal
distribution, then the corresponding Gaussian copula \(C_R\) is defined
through

\begin{equation}
\label{eq:gauss}
F_R(x_1, \ldots, x_d) = C_R(\Phi(x_1), \ldots, \Phi(x_d)),
\end{equation}

where \(\Phi(\cdot)\) is the standard normal CDF.

Here, the copula function \(C_R\) is the CDF of the random vector
\((\Phi(X_1), \ldots, \Phi(X_d))\), where
\((X_1, \ldots, X_d) \sim N_d({\bf 0}, R_P)\).

Sklar's Theorem (Úbeda-Flores and Fernández-Sánchez 2017) guarantees
that given inverse CDFs \(F_i^{-1}\)s and a valid correlation matrix
(within the Frechet bounds) a random vector can be obtained via
transformations involving copula functions. For example, using Gaussian
copulas, we can construct a random vector
\({\bf Y} = (Y_1, \ldots, Y_d)^\top\) with \(Y_i \sim F_i\),
viz.~\(Y_i = F_i^{-1}(\Phi(X_i)), i=1, \ldots, d\), where
\((X_1, \ldots, X_d) \sim N_d({\bf 0}, R_P)\).

\hypertarget{other-multivariate-simulation-packages}{%
\subsection{Other Multivariate Simulation
Packages}\label{other-multivariate-simulation-packages}}

To place \texttt{Bigsimr}'s implementation and algorithms in context of
existing approaches, we qualitatively compare \texttt{Bbigsimr} to other
multivariate simulation packages. In \texttt{R}, the \texttt{MASS}
package's \texttt{mvnorm()} has long been available to produce random
vectors from multivariate normal (and \(t\)) distributions. This
procedure cannot simulate from arbitrary margins and dependency measures
and it fails to readily scale to high dimensions. SAS PROC SIMNORMAL and
Python's NumPy \texttt{random.multivariate\_normal} both do similar
tasks. Recently, the high-performance \texttt{R} package
\texttt{mvnfast} (Fasiolo 2016) has been released, which provides HD
multivariate normal generation. However, lacks flexibility in marginal
models and dependency modeling. For dependent discrete data, LD
\texttt{R} packages exist, such as \texttt{GenOrd}. For copula
computation, the full-featured \texttt{R} package \texttt{copula}
provides many different copulas and a fairly general LD random generator
\texttt{rCopula()}. High-dimesional random vector generation via
\texttt{copula} takes an impractical amount of computing time (X. Li et
al. 2019). The \texttt{nortaRA} R package matches Pearson coefficients
near exactly and is reasonable for LD settings. Finally, we know of no
other \texttt{R} package that can produce HD random vectors with
flexible margins and dependence structure. Table \ref{tab:compare-table}
summarizes the above discussion.

\begin{table}[h]
\centering
\caption{\label{tab:compare-table}Comparison of selected features of bigsimr to existing R packages for multivariate simulation.}
\begin{tabular}{@{}llll@{}}
Package & Scales to HD & Hetereogenous Margins & Flexible Dependence \\ \midrule
MASS      & No           & No                    & No                  \\
mvnfast   & Yes          & No                    & No                  \\
GenOrd    & No           & Discrete margins      & Yes                 \\
copula    & No           & Yes                   & No                  \\
nortaRA   & No           & Yes                   & No                  \\
bigsimr   & Yes          & Yes                   & Yes                
\end{tabular}
\end{table}

\hypertarget{algorithms}{%
\section{Algorithms}\label{algorithms}}

This section describes our methods for simulating a random vector
\(\bf Y\) with \(Y_i\) components for \(i=1,\ldots,d\). Each \(Y_i\) has
a specified marginal CDF \(F_i\) and its inverse \(F^{-1}_i\). To
characterize dependency, every pair \((Y_i, Y_j)\) has a given Pearson
correlation \(\rho_P\), Spearman correlation \(\rho_S\), and/or
Kendall's \(\tau\). The method can be described as a
\emph{high-performance Gaussian copula} (Equation \ref{eq:gauss})
providing a HD NORTA-inspired algorithm.

\hypertarget{normal-to-anything-norta}{%
\subsection{NORmal To Anything (NORTA)}\label{normal-to-anything-norta}}

The well-known NORTA algorithm (Cario and Nelson 1997) simulates a
random vector \(\bf Y\) with variance-covariance matrix
\(\Sigma_{\bf Y}\). Specifically, NORTA algorithm proceeds as follows:

\begin{enumerate}
\def\labelenumi{\arabic{enumi}.}
\tightlist
\item
  Simulate a random vector \(\bf Z\) with \(d\) independent and
  identically distributed (iid) standard normal components.
\item
  Determine the input matrix \(\Sigma_{\bf Z}\) that corresponds with
  the specified output \(\Sigma_{\bf Y}\).
\item
  Produce a Cholesky factor \(M\) of \(\Sigma_{\bf Z}\) such that
  \(M M^{\prime}=\Sigma_{\bf Z}\).
\item
  Set \(X\) by \(X \gets MZ\).
\item
  \(\text{Return} \; {\bf Y} \; \text{where} \; Y_i \gets F_i^{-1}[\Phi(X_i)], \; i=1,...,d\).
\end{enumerate}

With modern parallel computing, steps 1, 3, 4, 5 are readily implemented
as high-performance, multi-core and/or graphical-processing-unit (GPU)
accelerated algorithms --- providing fast scalability to HD.

Matching specified Pearson correlation coefficients exactly (step 2
above), however, is computationally costly. In general, there is no
convenient correspondence between the components of the input
\(\Sigma_{\bf Z}\) and target \(\Sigma_{\bf Y}\). Matching the
correlations involves evaluating or approximating \(\binom{d}{2}\)
integrals of the form

\begin{equation}
    \mathrm{E}\left[Y_i Y_j\right] = \int_{-\infty}^{\infty} \int_{-\infty}^{\infty} F_i^{-1}\left[\Phi(z_i)\right] F_j^{-1}\left[\Phi(z_j)\right] \phi(z_i, z_j, \rho_z) dz_i dz_j,
    \label{eq:pearsonIntegralRelation}
\end{equation}

where \(\phi(\cdot)\) is the joint probability density function of two
correlated standard normal variables. For HD simulation, exact
evaluation may become too costly to enable practical simulation studies.
Our remedy to enable HD Pearson matching in \texttt{bigsimr} is Hermite
polynomial approximation of these \(\binom{d}{2}\) double integrals, as
suggested by Xiao and Zhou (2019).

\emph{NORTA in higher dimensions}

Sklar's theorem provides a useful characterization of multivariate
distributions through copulas. In practice, however, the particulars of
the simulation algorithm affect which joint distributions may be
simulated. Even in LD spaces (e.g., \(d=3\)), there exist valid
multivariate distributions with \emph{feasible} Pearson correlation
matrices that NORTA cannot match exactly (S. T. Li and Hammond 1975).
This occurs when the bivariate transformations are applied to find the
input correlation matrix, yet when combined, the resultant matrix
becomes non-positive definite. These situations do occur, even using
exact analytic calculations. Such problematic target correlation
matrices are termed \emph{NORTA defective}.

Ghosh and Henderson (2002) conducted a study to estimate the probability
of encountering NORTA defective matrices while increasing the dimension
\(d\). They found that for what is now considered low-to-moderate
dimensions (\(d \approx 20\)), almost \emph{all} feasible matrices are
NORTA defective. This stems from the concentration of measure near the
boundary of the space of all possible correlation matrices as dimension
increases. Unfortunately, it is precisely near this boundary that NORTA
defective matrices reside.

There is hope, however, as Ghosh and Henderson (2002) also showed that
replacing an non-positive definite input correlation matrix with a close
proxy will give approximate matching to the target --- with adequate
performance for moderate \(d\). This provides evidence that our nearest
positive definite (PD) augmented approach has promise to provide
adequate accuracy if our input matching scheme returns an indefinite
Pearson correlation matrix.

\hypertarget{rand-vec-gen}{%
\subsection{Random vector generator}\label{rand-vec-gen}}

We now describe our algorithm to generate random vectors, which extends
NORTA to HD via nearest correlation matrix replacement and provides
rank-based dependency matching:

\begin{enumerate}
\def\labelenumi{\arabic{enumi}.}
\tightlist
\item
  Mapping step, either

  \begin{itemize}
  \tightlist
  \item
    Convert the target Spearman correlation matrix \(R_S\) to the
    corresponding MVN Pearson correlation \(R_X\). Alternatively,
  \item
    Convert the target Kendall \(\tau\) matrix \(R_K\) to the
    corresponding MVN Pearson correlation \(R_X\). Alternatively,
  \item
    Convert the target Pearson correlation matrix to \(R_P\) to the
    corresponding, approximate MVN Pearson correlation \(R_X\).
  \end{itemize}
\item
  Check correlation matrix admissibility and, if not, compute the
  nearest correlation matrix.

  \begin{itemize}
  \tightlist
  \item
    Check that \(R_X\) is a correlation matrix, a positive definite
    matrix with 1's along the diagonal.
  \item
    If the mapping produced an \(R_X\) that is a correlation matrix, we
    describe the matrix as \textbf{admissible} in this scheme and retain
    \(R_X\). Otherwise,
  \item
    Replace \(R_X\) with the nearest correlation matrix \(\tilde{R}_X\),
    in the Frobenius norm and the result will be approximate.
  \end{itemize}
\item
  Gaussian copula

  \begin{itemize}
  \tightlist
  \item
    Generate \({\bf X}=(X_1, \ldots, X_d) \sim N_d({\bf 0}, R_X)\).
  \item
    Transform \({\bf X}\) to \({\bf U} = (U_1, \ldots, U_d)\)
    viz.~\(U_i=\Phi(X_i)\), \(i=1, \ldots, d\).
  \item
    Return \({\bf Y} = (Y_1, \ldots, Y_d)\), where
    \(Y_i=F_i^{-1}(U_i)\), \(i=1, \ldots, d\).
  \end{itemize}
\end{enumerate}

\emph{Step 1: Mapping step.} We first employ the closed-form
relationships between \(\rho_S\) and \(\tau\) with \(\rho_P\) for
bivariate normal random variables via Equations \ref{eq:convertKendall}
and \ref{eq:convertSpearman}, respectively (implemented as
\texttt{cor\_covert()}). Initializing our algorithm to match the
nonparametric correlations by computing these equations for all pairs is
computationally trivial.

For the computationally expensive process to match Pearson correlations,
we approximate Equation \ref{eq:pearsonIntegralRelation} for all pairs
of margins. To this end, we implement the approximation scheme
introduced by (Xiao and Zhou 2019). Briefly, the many double integrals
of the form in \ref{eq:pearsonIntegralRelation} are approximated by
weighted sums of Hermite polynomials. Matching coefficients for pairs of
continuous distributions is made tractable by this method, but for
discrete distributions (especially discrete distributions with large
support sets or infinite support), the approximation is computationally
expensive. To remedy this, we further approximate discrete distributions
by a continuous distribution using Generalized S-Distributions (Muino,
Voit, and Sorribas 2006).

\emph{Step 2: Admissibility check and nearest correlation matrix
computation.} Once \(R_X\) has been determined, we check admissibility
of the adjusted correlation via the steps described above. If \(R_X\) is
not a valid correlation matrix, then we compute the nearest correlation
matrix. Finding the nearest correlation matrix is a common statistical
computing problem. The defacto function in \texttt{R} is
\texttt{Matrix::nearPD}, an alternating projection algorithm due to
Higham (2002). As implemented, the function fails to scale to HD.
Instead, we provide the quadratically-convergent algorithm based on the
theory of strongly semi-smooth matrix functions (Qi and Sun 2006). The
nearest correlation matrix problem can be written down as the following
convex optimization problem:
\(\mathrm{min} \quad \frac{1}{2} \Vert R_X - X \Vert^2, \quad \mathrm{s.t.} \quad X_{ii} = 1, \quad i = 1, \ldots , n, \quad X \in S_{+}^{n}\).

For nonparametric correlation measures, our algorithm allows the
generation of HD multivariate data with arbitrary marginal distributions
with a broad class of admissible Spearman correlation matrices and
Kendall \(\tau\) matrices. The admissible classes consist of the
matrices that map to a Pearson correlation matrix for a MVN. In
particular, if we let \(X\) be MVN with \(d\) components and denote
\(\Omega_P = \{ R_P : R_P \textrm{ is a Pearson correlation matrix for } X \}, \quad \Omega_K = \{ R_K : R_K \textrm{ is a Kendall correlation matrix for } X \}, \quad \Omega_S = \{ R_S : R_S \textrm{ is a Spearman correlation matrix for } X \}\).
There are 1-1 mappings between these sets. We conjecture that the sets
of admissible \(R_S\) and \(R_K\) are not highly restrictive. In
particular, \(R_P\) is approximately \(R_S\) for a MVN, suggesting that
the admissible set \(\Omega_S\) should be flexible as \(R_P\) can be any
PD matrix with 1's along the diagonal. We provide methods to check
whether a target \(R_S\) is an element of the \emph{admissible set}
\(\Omega_S\).

There is an increasing probability of encountering an inadmissible
correlation matrix as dimension increases. In our experience, the
mapping step for large \(d\) almost always produces a \(R_X\) that is
not a correlation matrix. In Section \ref{package}, we provide a basic
method of how to quantify and control the approximation error. Further,
the RNA-seq example in Section \ref{examples} provides an illustration
of this in practice.

\emph{Step 3: Gaussian copula.} The final step implements a
NORTA-inspired, Gaussian copula approach to produce the desired margins.
Steps 1 and 2 determine the MVN Pearson correlation values that will
eventually match the target correlation. Step 3 requires a fast MVN
simulator, a standard normal CDF, and well-defined quantile functions
for marginals. The MVN is transformed to a copula (distribution with
standard uniform margins) by applying the normal CDF \(\Phi(\cdot)\).
Finally, the quantile functions \(F_i^{-1}\) are applied across the
margins to return the desired random vector \({\bf Y}\).

\hypertarget{the-bigsimr-r-package}{%
\section{The bigsimr R package}\label{the-bigsimr-r-package}}

This section describes a bivariate random vector simulation workflow via
the \texttt{bigsimr} \texttt{R} package. This \texttt{R} package
provides an interface to the native code written in Julia (registered as
the \texttt{Bigsimr} Julia package). In addition to the native Julia
\texttt{Bigsimr} package and \texttt{R} interface \texttt{bigsimr}, we
also provide a Python interface \texttt{bigsimr} that interfaces with
the Julia \texttt{Bigsimr} package. The Julia package provides a
high-performance implementation of our proposed random vector generation
algorithm and associated functions (see Section \ref{algorithms}).

The subsections below describe the basic use of the \texttt{bigsimr} R
package by stepping through an example workflow using the data set
\texttt{airquality} that contains daily air quality measurements in New
York, May to September 1973 (Chambers et al. 1983). This workflow
proceeds from setting up the computing environment, to data wrangling,
estimation, simulation configuration, random vector generation, and,
finally, result visualization.

\hypertarget{bivariate-example}{%
\subsection{Bivariate example}\label{bivariate-example}}

We illustrate the use of \texttt{bigsimr} using the New York air quality
data set (\texttt{airquality}) included in the R \texttt{datasets}
package. First, we load the \texttt{bigsimr} library and a few other
convenient data science packages, including the syntactically-elegant
\texttt{tidyverse} suite of \texttt{R} packages. The code chunk below
prepares the computing environment:

\begin{Shaded}
\begin{Highlighting}[]
\FunctionTok{library}\NormalTok{(}\StringTok{"tidyverse"}\NormalTok{)}
\FunctionTok{library}\NormalTok{(}\StringTok{"bigsimr"}\NormalTok{)}
\CommentTok{\# Activate multithreading in Julia}
\FunctionTok{Sys.setenv}\NormalTok{(}\AttributeTok{JULIA\_NUM\_THREADS =}\NormalTok{ parallel}\SpecialCharTok{::}\FunctionTok{detectCores}\NormalTok{())}
\CommentTok{\# Load the Bigsimr and Distributions Julia packages}
\NormalTok{bs }\OtherTok{\textless{}{-}} \FunctionTok{bigsimr\_setup}\NormalTok{()}
\NormalTok{dist }\OtherTok{\textless{}{-}} \FunctionTok{distributions\_setup}\NormalTok{()}
\end{Highlighting}
\end{Shaded}

Here, we describe a minimal working example --- a bivariate simulation
of two airquality variables: \texttt{Temperature}, in degrees
Fahrenheit, and \texttt{Ozone} level, in parts per billion.

\begin{Shaded}
\begin{Highlighting}[]
\NormalTok{df }\OtherTok{\textless{}{-}}\NormalTok{ airquality }\SpecialCharTok{\%\textgreater{}\%} \FunctionTok{select}\NormalTok{(Temp, Ozone) }\SpecialCharTok{\%\textgreater{}\%} \FunctionTok{drop\_na}\NormalTok{()}
\FunctionTok{glimpse}\NormalTok{(df)}
\end{Highlighting}
\end{Shaded}

\begin{verbatim}
## Rows: 116
## Columns: 2
## $ Temp  <int> 67, 72, 74, 62, 66, 65, 59, 61, 74, 69, 66, 68, 58, 64, 66, 5...
## $ Ozone <int> 41, 36, 12, 18, 28, 23, 19, 8, 7, 16, 11, 14, 18, 14, 34, 6, ...
\end{verbatim}

\begin{figure}
\centering
\includegraphics{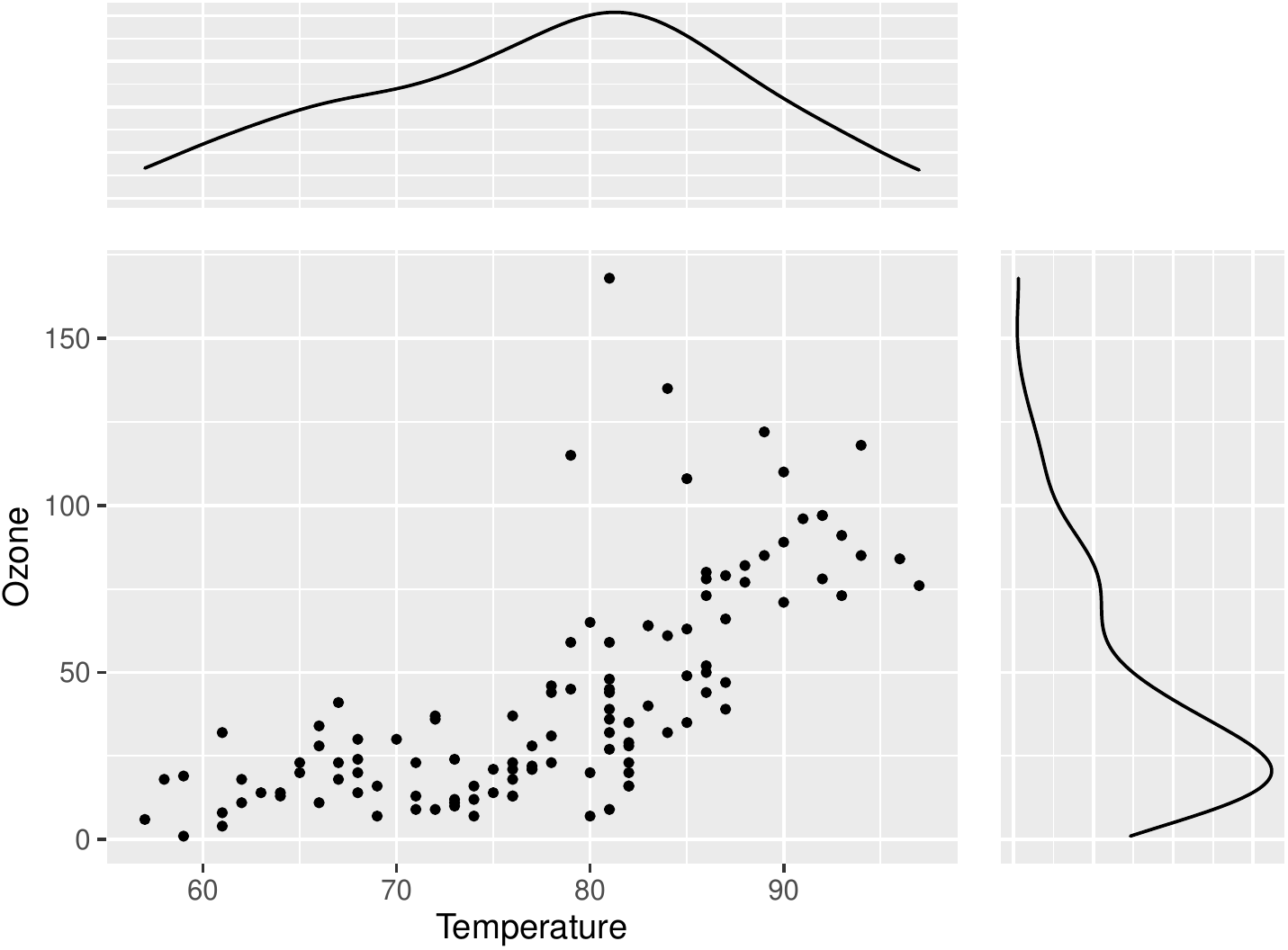}
\caption{\label{fig:ch030-aq-joint-dist}Bivariate scatterplot of Ozone
versus Temp with estimated marginal densities. The Ozone data are
modeled marginally as log-normal and the Temperature data as normal.}
\end{figure}

Figure \ref{fig:ch030-aq-joint-dist} visualizes the bivariate
relationship between \texttt{Ozone} and \texttt{Temperature}. We aim to
simulate random two-component vectors mimicking this structure. The
margins are not normally distributed; \texttt{Ozone} level exhibits a
strong positive skew.

Next, we specify the marginal distributions and correlation coefficient
(both type and magnitude). Here the analyst is free to be creative. For
this example, we avoid goodness-of-fit considerations to determine the
marginal distributions. It is sensible without domain knowledge to
estimate these quantities from the data, and \texttt{bigsimr} contains
fast functions designed for this task.

\hypertarget{specifying-marginal-distributions}{%
\subsection{Specifying marginal
distributions}\label{specifying-marginal-distributions}}

Based on the estimated densities in Figure
\ref{fig:ch030-aq-joint-dist}, we assume \texttt{Temp} is normally
distributed and \texttt{Ozone} is log-normally distributed, as the
latter values are positive and skewed. We use the classical unbiased
estimators for the normal distribution's parameters and maximum
likelihood estimators for the log-normal parameters:

\begin{Shaded}
\begin{Highlighting}[]
\NormalTok{df }\SpecialCharTok{\%\textgreater{}\%} \FunctionTok{select}\NormalTok{(Temp) }\SpecialCharTok{\%\textgreater{}\%} 
  \FunctionTok{summarise\_all}\NormalTok{(}\AttributeTok{.funs =} \FunctionTok{c}\NormalTok{(}\AttributeTok{mean =}\NormalTok{ mean, }\AttributeTok{sd =}\NormalTok{ sd))}
\end{Highlighting}
\end{Shaded}

\begin{verbatim}
##       mean       sd
## 1 77.87069 9.485486
\end{verbatim}

\begin{Shaded}
\begin{Highlighting}[]
\NormalTok{mle\_mean }\OtherTok{\textless{}{-}} \ControlFlowTok{function}\NormalTok{(x) }\FunctionTok{mean}\NormalTok{(}\FunctionTok{log}\NormalTok{(x))}
\NormalTok{mle\_sd }\OtherTok{\textless{}{-}} \ControlFlowTok{function}\NormalTok{(x) }\FunctionTok{mean}\NormalTok{( }\FunctionTok{sqrt}\NormalTok{( (}\FunctionTok{log}\NormalTok{(x) }\SpecialCharTok{{-}} \FunctionTok{mean}\NormalTok{(}\FunctionTok{log}\NormalTok{(x)))}\SpecialCharTok{\^{}}\DecValTok{2}\NormalTok{ ) )}
\NormalTok{df }\SpecialCharTok{\%\textgreater{}\%} 
  \FunctionTok{select}\NormalTok{(Ozone) }\SpecialCharTok{\%\textgreater{}\%} 
  \FunctionTok{summarise\_all}\NormalTok{(}\AttributeTok{.funs =} \FunctionTok{c}\NormalTok{(}\AttributeTok{meanlog =}\NormalTok{ mle\_mean, }\AttributeTok{sdlog =}\NormalTok{ mle\_sd))}
\end{Highlighting}
\end{Shaded}

\begin{verbatim}
##    meanlog     sdlog
## 1 3.418515 0.6966689
\end{verbatim}

Now, we configure the input marginals for later input into
\texttt{rvec}. The marginal distributions are specified using
\texttt{Julia}'s \texttt{Distributions} package and stored in a vector.

\begin{Shaded}
\begin{Highlighting}[]
\NormalTok{margins }\OtherTok{\textless{}{-}} \FunctionTok{c}\NormalTok{(dist}\SpecialCharTok{$}\FunctionTok{Normal}\NormalTok{(}\FunctionTok{mean}\NormalTok{(df}\SpecialCharTok{$}\NormalTok{Temp), }\FunctionTok{sd}\NormalTok{(df}\SpecialCharTok{$}\NormalTok{Temp)),}
\NormalTok{             dist}\SpecialCharTok{$}\FunctionTok{LogNormal}\NormalTok{(}\FunctionTok{mle\_mean}\NormalTok{(df}\SpecialCharTok{$}\NormalTok{Ozone), }\FunctionTok{mle\_sd}\NormalTok{(df}\SpecialCharTok{$}\NormalTok{Ozone)))}
\end{Highlighting}
\end{Shaded}

\hypertarget{specifying-correlation}{%
\subsection{Specifying correlation}\label{specifying-correlation}}

The user must decide how to describe correlation based on the
particulars of the problem. For non-normal data in our scheme, we
advocate the use of Spearman's \(\rho\) correlation matrix \(R_S\) or
Kendall's \(\tau\) correlation matrix \(R_K\). We also support Pearson
correlation coefficient matching, while cautioning the user to check the
performance for the distribution at hand (see
\protect\hyperlink{simulations}{Monte Carlo evaluations} below for
evaluation strategies and guidance). Note that these estimation methods
are classical approaches, not specifically designed for high-dimensional
correlation estimation.

\begin{Shaded}
\begin{Highlighting}[]
\NormalTok{(R\_S }\OtherTok{\textless{}{-}}\NormalTok{ bs}\SpecialCharTok{$}\FunctionTok{cor}\NormalTok{(}\FunctionTok{as.matrix}\NormalTok{(df), bs}\SpecialCharTok{$}\NormalTok{Spearman))}
\end{Highlighting}
\end{Shaded}

\begin{verbatim}
##          [,1]     [,2]
## [1,] 1.000000 0.774043
## [2,] 0.774043 1.000000
\end{verbatim}

\hypertarget{checking-target-correlation-matrix-admissibility}{%
\subsection{Checking target correlation matrix
admissibility}\label{checking-target-correlation-matrix-admissibility}}

Once a target correlation matrix is specified, first convert the
dependency type to Pearson to correctly specify the MVN inputs. Then,
check that the converted matrix \(R_X\) is a valid correlation matrix
(PD and 1s on the diagonal). For this bivariate example, \(R_X\) is a
valid correlation matrix. Note that typically in HD the resultant MVN
Pearson correlation matrix is indefinite and requires approximation (see
the HD examples in subsequent sections).

\begin{Shaded}
\begin{Highlighting}[]
\CommentTok{\# Step 1. Mapping}
\NormalTok{(R\_X }\OtherTok{\textless{}{-}}\NormalTok{ bs}\SpecialCharTok{$}\FunctionTok{cor\_convert}\NormalTok{(R\_S, bs}\SpecialCharTok{$}\NormalTok{Spearman, bs}\SpecialCharTok{$}\NormalTok{Pearson))}
\end{Highlighting}
\end{Shaded}

\begin{verbatim}
##           [,1]      [,2]
## [1,] 1.0000000 0.7885668
## [2,] 0.7885668 1.0000000
\end{verbatim}

\begin{Shaded}
\begin{Highlighting}[]
\CommentTok{\# Step 2. Check admissibility}
\NormalTok{bs}\SpecialCharTok{$}\FunctionTok{iscorrelation}\NormalTok{(R\_X)}
\end{Highlighting}
\end{Shaded}

\begin{verbatim}
## [1] TRUE
\end{verbatim}

Despite being a valid correlation matrix under this definition, marginal
distributions induce Frechet bounds on the possible Pearson correlation
values. To check that the converted correlation values fall within these
bounds, use \texttt{bigsimr::cor\_bounds} to estimate the pairwise lower
and upper correlation bounds. \texttt{cor\_bounds} uses the Generate,
Sort, and Correlate algorithm of Demirtas and Hedeker (2011). For the
assumed marginals, the Pearson correlation coefficient is not free to
vary in {[}-1,1{]}, as seen below. Our MC estimate of the bounds
slightly underestimates the theoretical bounds of \((-0.881, 0.881)\).
(See an analytic derivation presented in the Appendix). Since our single
Pearson correlation coefficient is within the theoretical bounds, the
correlation matrix is valid for our simulation strategy.

\begin{Shaded}
\begin{Highlighting}[]
\NormalTok{bs}\SpecialCharTok{$}\FunctionTok{cor\_bounds}\NormalTok{(margins[}\DecValTok{1}\NormalTok{], margins[}\DecValTok{2}\NormalTok{], bs}\SpecialCharTok{$}\NormalTok{Pearson, }\AttributeTok{n\_samples =} \FloatTok{1e6}\NormalTok{)}
\end{Highlighting}
\end{Shaded}

\begin{verbatim}
## Julia Object of type NamedTuple{(:lower, :upper),Tuple{Float64,Float64}}.
## (lower = -0.8829854580269612, upper = 0.882607249144551)
\end{verbatim}

In this 2D example, it is possible to verify the existence of a
bivariate distribution with the target Pearson correlation / margins
(See the {[}Appendix{]}). For general \(d-\)variate distributions, is it
not guaranteed that such a multivariate distribution exists with exact
the specified correlations and margins (Barbiero and Ferrari 2017).
Theoretical guidance is lacking in this regard, but, in practice, we
find that using bivariate bounds form rough approximations to the bounds
for the feasible region for the \(d-\)variate construction when
aggregated into a Pearson matrix. See the
\protect\hyperlink{examples}{RNA-seq data application} section for an
example discussing this, NORTA feasibility, and use of the nearest PD
matrix algorithm.

\hypertarget{simulating-random-vectors}{%
\subsection{Simulating random vectors}\label{simulating-random-vectors}}

\begin{figure}
\centering
\includegraphics{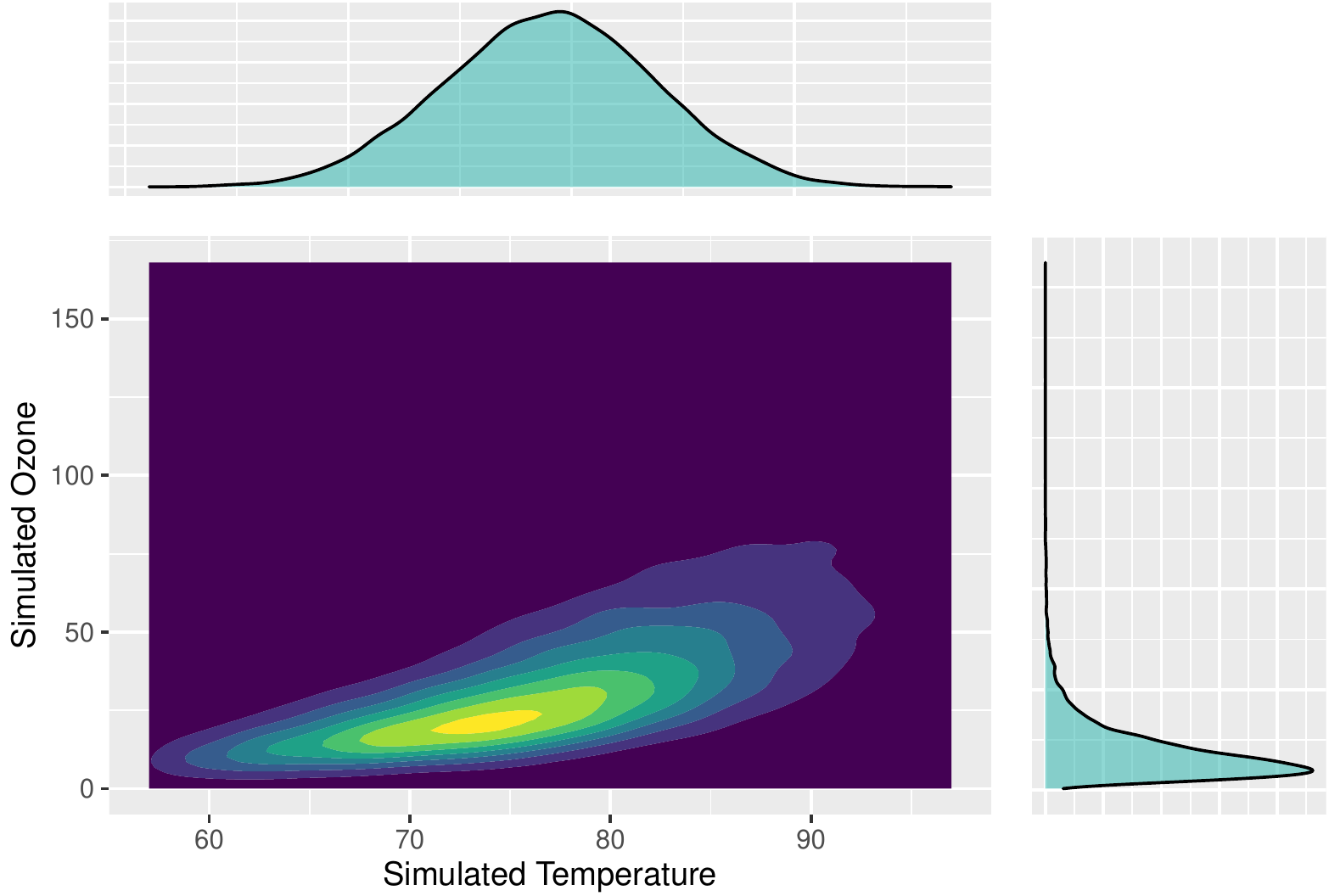}
\caption{\label{fig:ch030-plot-sim}Contour plot and marginal densities
for the simulated bivariate distribution of Air Quality Temperatures and
Ozone levels. The simulated points mimic the observed data with respect
to both the marginal characteristics and bivariate association.}
\end{figure}

Finally, we execute \texttt{rvec} to simulate the desired \(10,000\)
random vectors from the assumed bivariate distribution of \texttt{Ozone}
and \texttt{Temp}. Note that the target input is specified using the
pre-computed MVN correlation matrix. Figure \ref{fig:ch030-plot-sim}
plots the 10,000 simulated points.

\begin{Shaded}
\begin{Highlighting}[]
\NormalTok{x }\OtherTok{\textless{}{-}}\NormalTok{ bs}\SpecialCharTok{$}\FunctionTok{rvec}\NormalTok{(}\DecValTok{10000}\NormalTok{, R\_X, margins)}
\NormalTok{df\_sim }\OtherTok{\textless{}{-}} \FunctionTok{as.data.frame}\NormalTok{(x)}
\FunctionTok{colnames}\NormalTok{(df\_sim) }\OtherTok{\textless{}{-}} \FunctionTok{colnames}\NormalTok{(df)}
\end{Highlighting}
\end{Shaded}

\hypertarget{simulations}{%
\section{Monte Carlo evaluations}\label{simulations}}

In this section, we conduct MC studies to investigate method
performance. Marginal parameter matching is straightforward in our
scheme as it is a sequence of univariate inverse probability transforms.
This process is both fast and exact. On the other hand, accurate and
computationally efficient dependency matching presents a challenge,
especially in the HD setting. To evaluate our methods in those respects,
we design the following numerical experiments to first assess accuracy
of matching dependency parameters in bivariate simulations and then time
the procedure in increasingly large dimension \(d\).

\hypertarget{bivariate-experiments}{%
\subsection{Bivariate experiments}\label{bivariate-experiments}}

We select bivariate simulation configurations to ultimately simulate our
motivating discrete-valued RNA-seq example. Thus, we proceed by
increasing departure from normality, leading to a multivariate negative
binomial (MVNB) model for our motivating data. We begin with empirically
evaluating the dependency matching across all three supported
correlations --- Pearson, Spearman, and Kendall --- in identical,
bivariate marginal configurations. For each pair of identical margins,
we vary the target correlation across \(\Omega\), the set of possible
admissible values for each correlation type, to evaluate the
simulation's ability to generate all possible correlations. The
simulations progress from bivariate normal, to bivariate gamma
(non-normal yet continuous), and bivariate negative binomial (non-normal
and discrete).

Table \ref{tab:sims} lists our identical-marginal, bivariate simulation
configurations. We increase the simulation replicates \(B\) to visually
ensure that our results converge to the target correlations and gauge
efficiency. We select distributions beginning with a standard
multivariate normal (MVN) as we expect the performance to be exact (up
to MC error) for all correlation types. Then, we select a non-symmetric
continuous distribution: a standard (rate =1), two-component
multivariate gamma (MVG). Finally, we select distributions and marginal
parameter values that are motivated by our RNA-seq data, namely values
proximal to probabilities and sizes estimated from the data (see
\href{examples}{RNA-seq data application} for estimation details).
Specifically, we assume a MVNB with
\(p_1 = p_2 = 3\times10^{-4}, r_1 = r_2 = 4, \rho \in \Omega\).

\begin{table}[]
\centering
\caption{ \label{tab:sims} Identical margin, bivariate simulation configurations to evaluate matching accuracy.}
\begin{tabular}{@{}lcr@{}}
\toprule
Simulation Reps $B$ & Correlation Types & Identical-margin 2D distribution \\ \midrule
$1,000$ & Pearson ($\rho_P$) & ${\bf Y} \sim MVN( \mu= 0 , \sigma = 1, \rho_i ), i=1,\ldots,100$ \\
$10,000$ & Spearman ($\rho_S$) & ${\bf Y} \sim MVG( shape = 10, rate = 1, \rho_i ), i=1,\ldots,100$ \\
$100,000$ & Kendall ($\tau$) & ${\bf Y} \sim MVNB(p = 3\times10^{-4}, r = 4,\rho_i), i=1,\ldots,100$ \\ \bottomrule
\end{tabular}
\end{table}

For each of the unique 9 simulation configurations described above, we
estimate the correlation bounds and vary the correlations along a
sequence of 100 points evenly placed within the bounds, aiming to
explore \(\Omega\). Specifically, we set correlations
\(\{ \rho_1 = ( \hat{l} + \epsilon), \rho_2 = (\hat{l} + \epsilon) + \delta, \ldots, \rho_{100} = (\hat{u} - \epsilon) \}\),
with \(\hat{l}\) and \(\hat{u}\) being the estimated lower and upper
bounds, respectively, with increment value \(\delta\). The adjustment
factor, \(\epsilon=0.01\), is introduced to handle numeric issues when
the bound is specified exactly.

\begin{figure}
\centering
\includegraphics{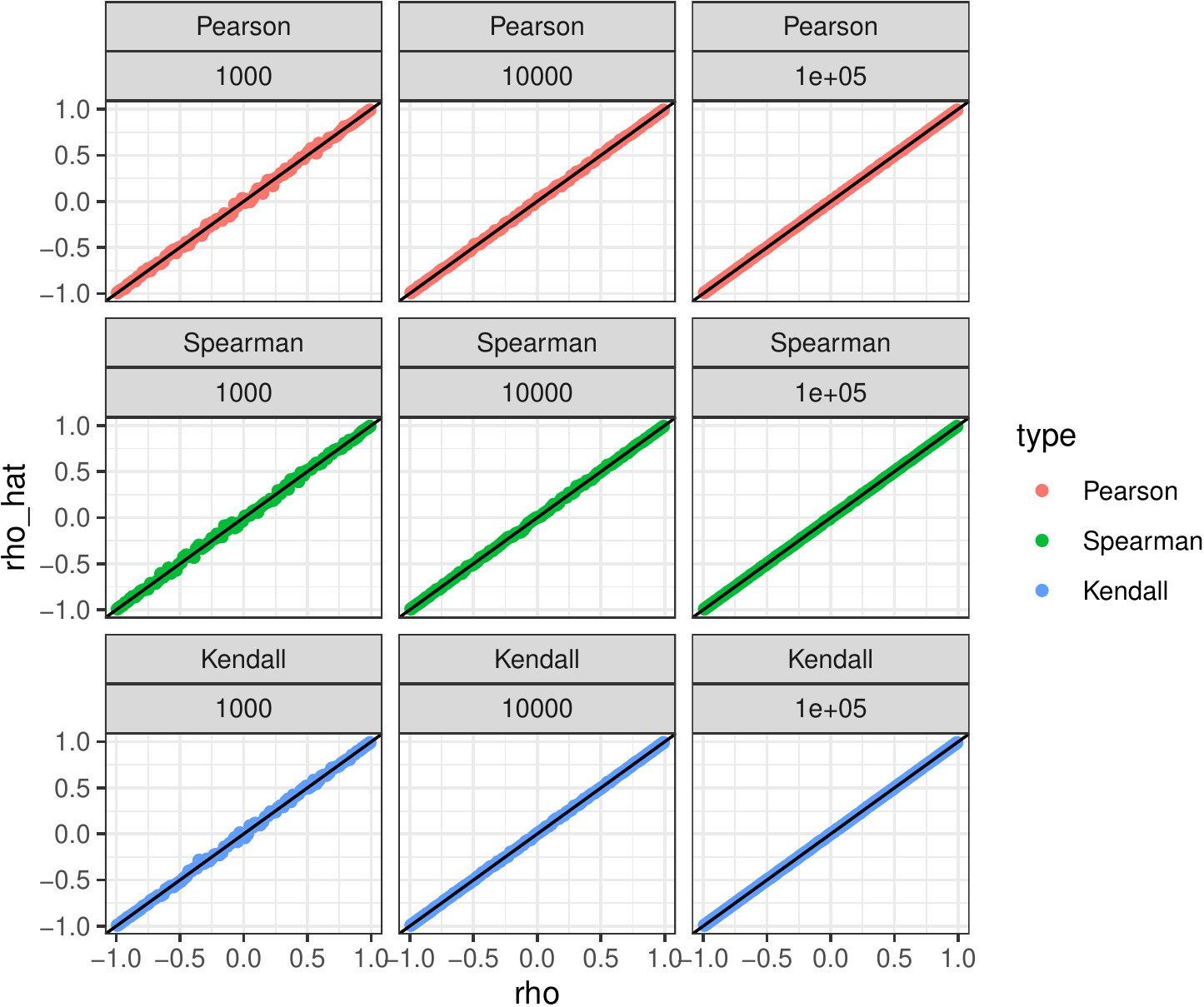}
\caption{\label{fig:ch040-biNormPlot}\texttt{bigsimr} recovers the
Pearson specified correlations for MVN.}
\end{figure}

\begin{figure}
\centering
\includegraphics{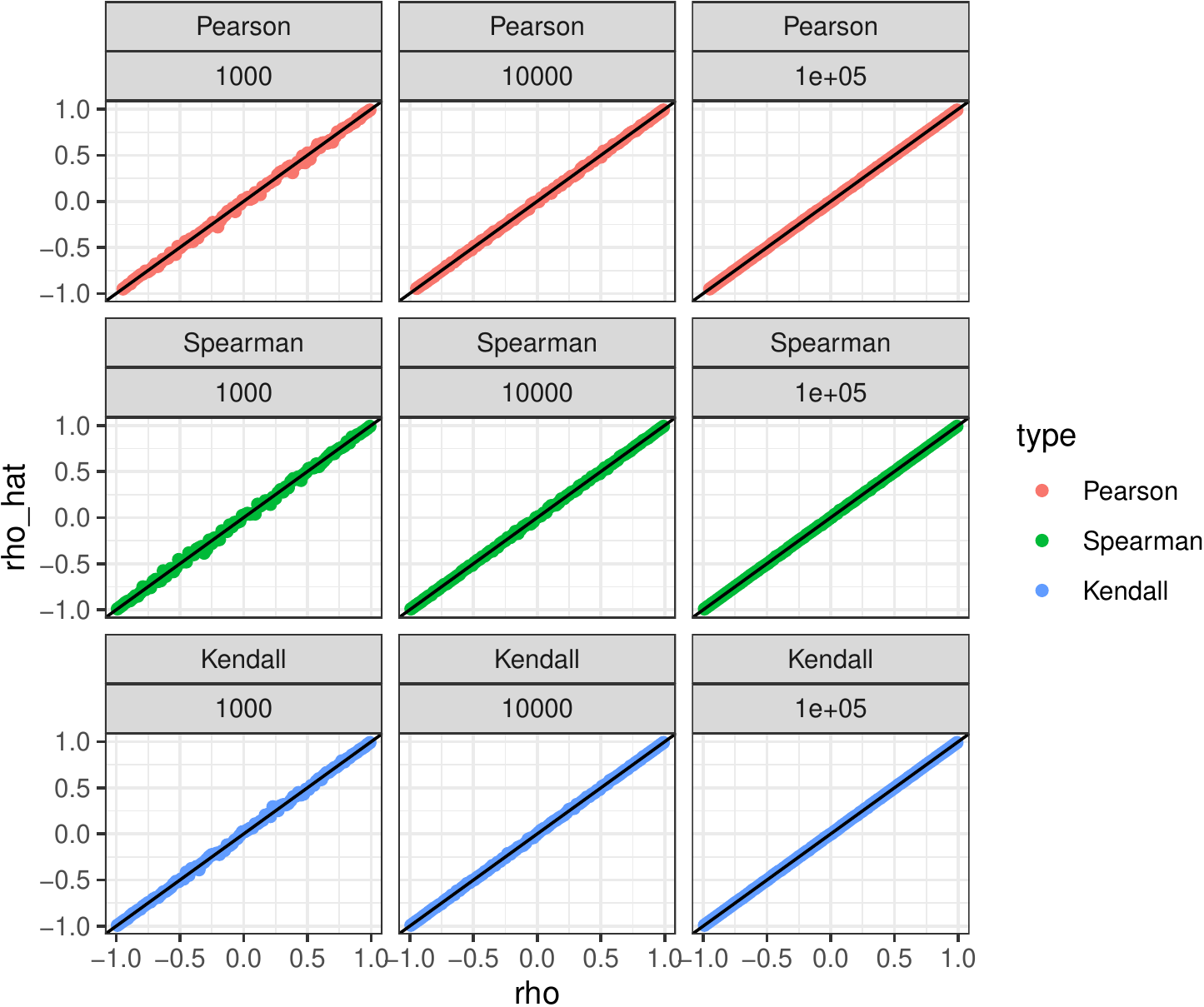}
\caption{\label{fig:ch040-biGammaPlot}\texttt{bigsimr} recovers the
Pearson specified correlations for Bivariate Gamma.}
\end{figure}

\begin{figure}
\centering
\includegraphics{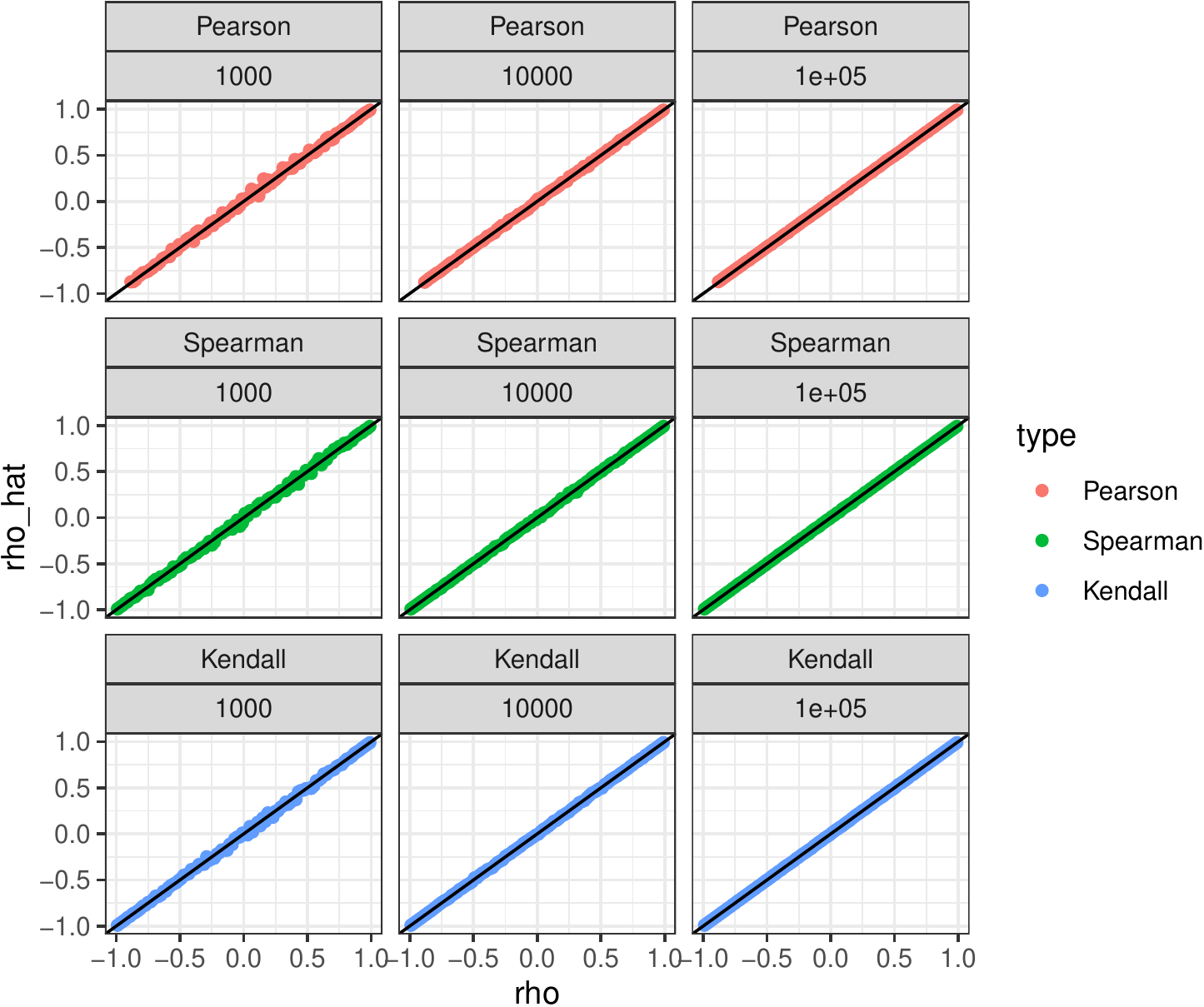}
\caption{\label{fig:ch040-biNegBinPlot}\texttt{bigsimr} recovers the
correlations for bivariate negative binomial only approximately for
Pearson but (nearly) exactly for the rank-based correlations.}
\end{figure}

\begin{figure}
\centering
\includegraphics{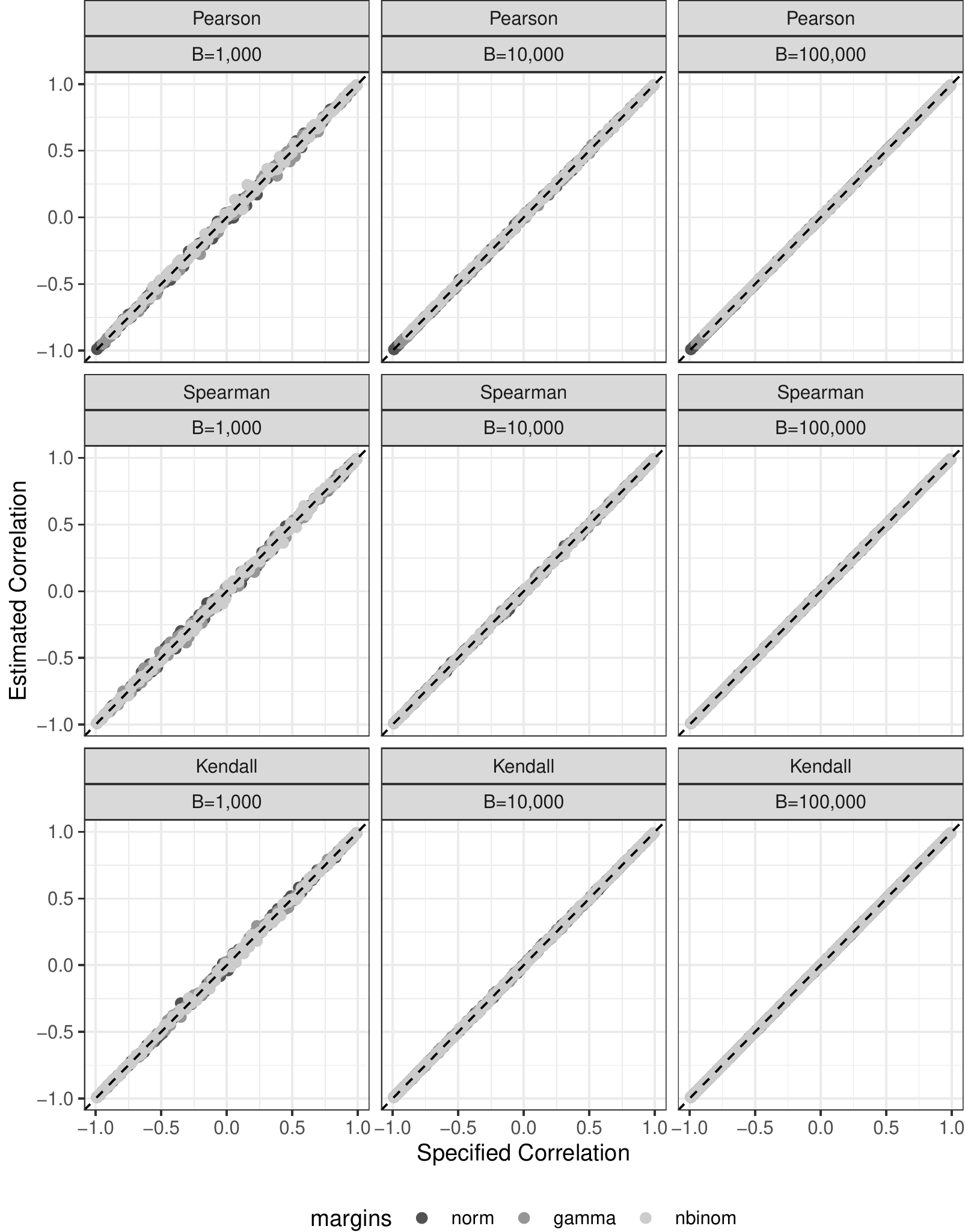}
\caption{\label{fig:ch040-bPlot}Bivariate simulations match target
correlations across the entire range of feasible correlations. The
horizontal axis plots the specified target correlations for each
bivariate margin. Normal margins are plotted in dark dark grey, gamma in
medium grey, and negative binomial in light grey. As the number of
simulated vectors \(B\) increases from left to right, the variation in
estimated correlations (vertical axis) decreases. The dashed line
indicates equality between the specified and estimated correlations.}
\end{figure}

Figure \ref{fig:ch040-bPlot} displays the aggregated bivariate
simulation results. Table \ref{tab:ch040-BiError} contains the mean
absolute error (MAE) in reproducing the desired dependency measures for
the three bivariate scenarios. Overall, the studies show that our
methodology is generally accurate across the entire range of possible
correlation for all three dependency measures. Our Pearson matching
performs nearly as well as Spearman or Kendall, except for a slight
increase in error for the negative binomial case.

\begin{table}

\caption{\label{tab:ch040-BiError}Average abolute error in matching the target dependency across the entire range of possible correlations for each bivariate marginal.}
\centering
\begin{tabular}[t]{lllr}
\toprule
No. of random vectors & Correlation type & Distribution & Mean abs. error\\
\midrule
1000 & Pearson & norm & 0.0156184\\
1000 & Pearson & gamma & 0.0159287\\
1000 & Pearson & nbinom & 0.0162465\\
\addlinespace
1000 & Spearman & norm & 0.0175440\\
1000 & Spearman & gamma & 0.0182176\\
1000 & Spearman & nbinom & 0.0151124\\
\addlinespace
1000 & Kendall & norm & 0.0130284\\
1000 & Kendall & gamma & 0.0114740\\
1000 & Kendall & nbinom & 0.0126014\\
\addlinespace
10000 & Pearson & norm & 0.0060226\\
10000 & Pearson & gamma & 0.0057667\\
10000 & Pearson & nbinom & 0.0058180\\
\addlinespace
10000 & Spearman & norm & 0.0062388\\
10000 & Spearman & gamma & 0.0056132\\
10000 & Spearman & nbinom & 0.0049256\\
\addlinespace
10000 & Kendall & norm & 0.0032618\\
10000 & Kendall & gamma & 0.0038067\\
10000 & Kendall & nbinom & 0.0033203\\
\addlinespace
1e+05 & Pearson & norm & 0.0017570\\
1e+05 & Pearson & gamma & 0.0016962\\
1e+05 & Pearson & nbinom & 0.0029582\\
\addlinespace
1e+05 & Spearman & norm & 0.0016607\\
1e+05 & Spearman & gamma & 0.0016408\\
1e+05 & Spearman & nbinom & 0.0015269\\
\addlinespace
1e+05 & Kendall & norm & 0.0010441\\
1e+05 & Kendall & gamma & 0.0011077\\
1e+05 & Kendall & nbinom & 0.0011976\\
\bottomrule
\end{tabular}
\end{table}

\hypertarget{scale-up-to-high-dimensions}{%
\subsection{Scale up to High
Dimensions}\label{scale-up-to-high-dimensions}}

With information of our method's accuracy from a low-dimensional
perspective, we now assess whether \texttt{bigsimr} can scale to larger
dimensional problems with practical computation times. We ultimately
generate \(B=1,000\) random vectors for
\(d=\{100, 250, 500, 1000, 2500, 5000, 10000\}\) for each correlation
type, \{Pearson, Spearman, Kendall\} while timing the algorithm's major
steps. We begin by producing a synthetic ``data set'' by completing the
following steps:

\begin{enumerate}
\def\labelenumi{\arabic{enumi}.}
\tightlist
\item
  Produce heterogeneous gamma marginals by randomly selecting the
  \(j^{th}\) gamma shape parameter from
  \(U_j \sim uniform(1,10), j=1,\ldots,d\) and the \(j^{th}\) rate
  parameter from \(V_j \sim exp(1/5), j=1,\ldots,d\), with the constant
  parameters determined arbitrarily.
\item
  Produce a random full-rank Pearson correlation matrix via
  \texttt{cor\_randPD} of size \(d \times d\).
\item
  Simulate a ``data set'' of \(1,000 \times d\) random vectors via
  \texttt{rvec}.
\end{enumerate}

With the synthetic data set in hand, we complete and time the following
four steps involved in a typical workflow (including correlation
estimation).

\begin{enumerate}
\def\labelenumi{\arabic{enumi}.}
\tightlist
\item
  Estimate the correlation matrix from the ``data'' in the \emph{Compute
  Correlation} step.
\item
  Map the correlations to initialize the algorithm (Pearson to Pearson,
  Spearman to Pearson, or Kendall to Pearson) in the \emph{Adjust
  Correlation} step.
\item
  Check whether the mapping produces a valid correlation matrix and, if
  not, find the nearest PD correlation matrix in the \emph{Check
  Admissibility} step.
\item
  Simulate \(1,000\) vectors in the \emph{Simulate Data} step.
\end{enumerate}

The experiments are conducted on a MacBook Pro carrying a 2.4 GHz 8-Core
Intel Core i9 processor, with all 16 threads employed during
computation. Table \ref{tab:ch040-moderateDtab} displays the total
computation time for moderate dimesions (\(d \leq 500\)). In every
simullation setting, the 1,000 vectors are generated rapidly, executing
in under 2 seconds.

\begin{table}

\caption{\label{tab:ch040-moderateDtab}Total time to produce 1,000 random vectors with a random correlation matrix and hetereogeneous gamma margins.}
\centering
\begin{tabular}[t]{llr}
\toprule
Dimension & Correlation type & Total Time (Seconds)\\
\midrule
100 & Pearson & 0.081\\
100 & Spearman & 0.027\\
100 & Kendall & 0.075\\
\addlinespace
250 & Pearson & 0.378\\
250 & Spearman & 0.062\\
250 & Kendall & 0.375\\
\addlinespace
500 & Pearson & 1.425\\
500 & Spearman & 0.135\\
500 & Kendall & 1.495\\
\bottomrule
\end{tabular}
\end{table}

\begin{figure}
\centering
\includegraphics{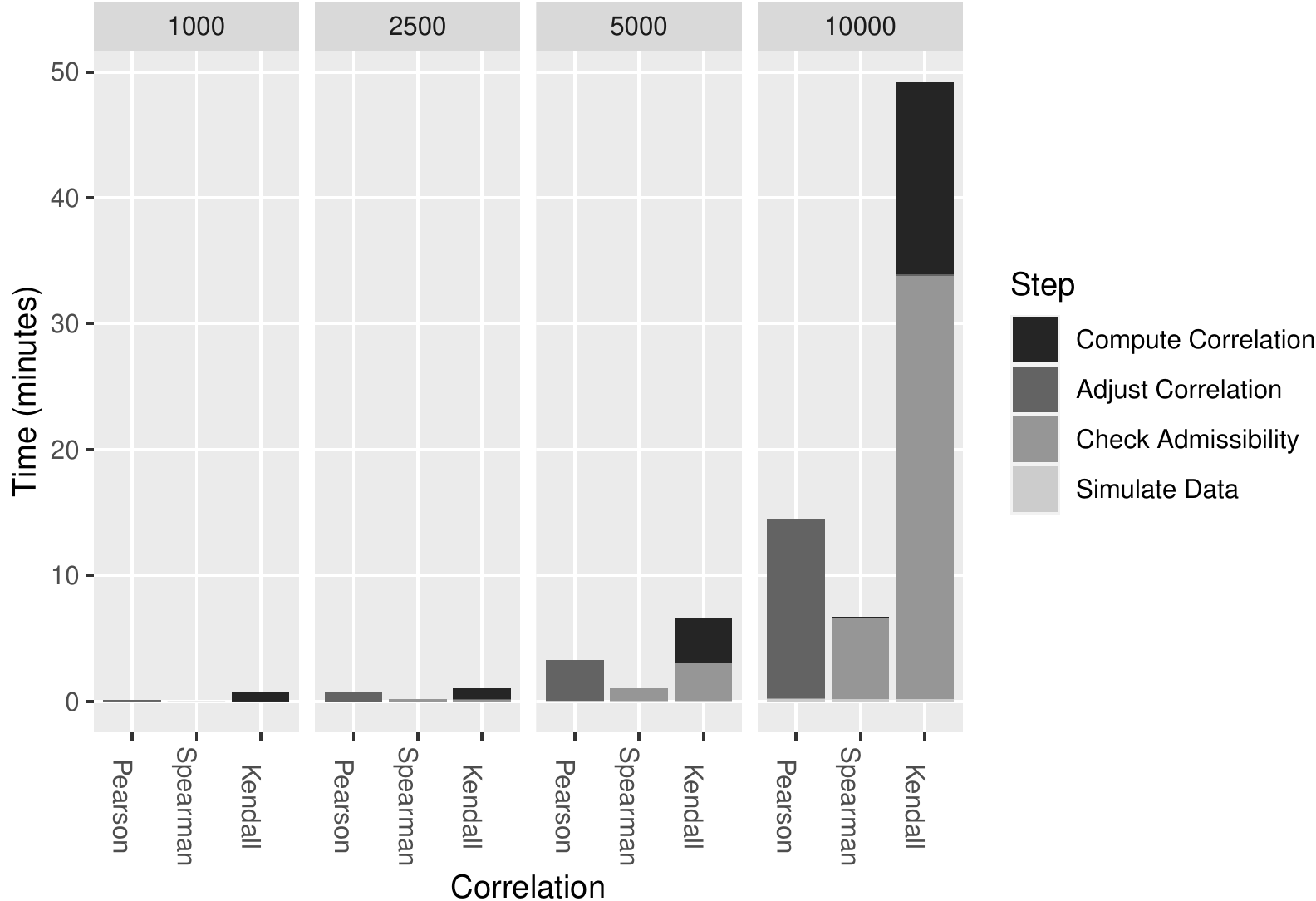}
\caption{\label{fig:ch040-largeDfig}Computation times for HD
multivariate gamma simulation.}
\end{figure}

The results for \(d > 500\) show scalability to ultra-high dimensions
for all three correlation types, although the total times do become much
larger. Figure \ref{fig:ch040-largeDfig} displays computation times for
\(d=\{1000, 2500, 5000, 10000\}\). For \(d\) equal to 1000 and 2500, the
total time is under a couple of minutes. At \(d\) of 5000 and 10,000,
Pearson correlation matching in the \emph{Adjust Correlation} step
becomes costly. Interestingly, Pearson is actually faster than Kendall
for \(d=10,000\) due to bottlenecks in \emph{Compute Correlation} and
\emph{Check Admissibility}. Uniformly, matching Spearman correlations is
faster, with total times under 5 minutes for \(d=10,000\), making
Spearman the most computationally-friendly dependency type for our
package. With this in mind, we scaled the simulation to \(d=20,000\) for
the Spearman type and obtained the 1,000 vectors in under an hour (data
not shown). In principle, this would enable the simulation of an entire
human-derived RNA-seq data set. We note that for a given target
correlation matrix and margins, steps 1, 2, and 3 only need to be
computed once and the fourth step, \emph{Simulate Data}, is nearly
instaneous for all settings considered.

\emph{Limitations, conclusions, and recommendations}

In the bivariate studies, we chose arbitrary simulation parameters for
three distributions, moving from the Gaussian to discrete and non-normal
MVNB. Under these conditions, the simulated random vectors sample the
desired bivariate distribution across the entire range of pairwise
correlations for the three dependency measures. The simulation results
could differ for other choices of simulation settings. Specifying
extreme correlations near the boundary or Frechet bounds could result in
poor simulation performance. Fortunately, it is straightforward to
evaluate simulation performance by using strategies similar to those
completed above. We expect our random vector generation to perform well
for the vast majority of NORTA-feasible correlation matrices, but advise
to check the performance before making inferences/further analyses.

Somewhat surprisingly, Kendall estimation and nearest PD computation
scale poorly compared to Spearman and, even, approximate Pearson
matching. In our experience, Kendall computation times are sensitive to
the number of cores, benefiting from multi-core parallelization. This
could mitigate some of the current algorithmic/implementation
shortcomings. Despite this, Kendall matching is still feasible for most
HD data sets. Finally, we note that one could use our single-pass
algorithm \texttt{cor\_fastPD} to produce a `close' (not nearest PD) to
scale to even higher dimensions with some loss of accuracy.

\hypertarget{examples}{%
\section{RNA-seq data application}\label{examples}}

This section demonstrates how to simulate multivariate data using
\texttt{bigsimr}, aiming to replicate the structure of HD dependent
count data. In an illustration of our proposed methodology, we seek to
simulate RNA-sequencing data by producing simulated random vectors
mimicking the observed data and its generating process. Modeling RNA-seq
using multivariate probability distributions is natural as inter-gene
correlation occurs during biological processes (Wang, Gerstein, and
Snyder 2009). And yet, many models do not account for this, leading to
major disruptions to the operating characteristics of statistical
estimation, testing, and prediction. The following subsections apply
\texttt{bigsimr}'s methods to real RNA-seq data, including replicating
an estimated parametric structure, probability estimation, and
evaluation of correlation estimation efficiency.

\hypertarget{simulating-high-dimensional-rna-seq-data}{%
\subsection{Simulating High-Dimensional RNA-seq
data}\label{simulating-high-dimensional-rna-seq-data}}

\begin{Shaded}
\begin{Highlighting}[]
\NormalTok{d }\OtherTok{\textless{}{-}} \DecValTok{1000}
\NormalTok{brca1000 }\OtherTok{\textless{}{-}}\NormalTok{ example\_brca }\SpecialCharTok{\%\textgreater{}\%}
  \FunctionTok{select}\NormalTok{(}\FunctionTok{all\_of}\NormalTok{(}\DecValTok{1}\SpecialCharTok{:}\NormalTok{d)) }\SpecialCharTok{\%\textgreater{}\%}
  \FunctionTok{mutate}\NormalTok{(}\FunctionTok{across}\NormalTok{(}\FunctionTok{everything}\NormalTok{(), as.double))}
\end{Highlighting}
\end{Shaded}

We begin by estimating the structure of the TCGA BRCA RNA-seq data set.
Ultimately, we will simulate \(B=10,000\) random vectors
\({\bf Y}=(Y_1, \ldots, Y_d)^\top\) with \(d=1000\). We assume a MVNB
model as RNA-seq counts are often over-dispersed and correlated. Since
all \(d\) selected genes exhibit over-dispersion (data not shown), we
proceed to estimate the NB parameters \((r_i, p_i), i=1,\ldots,d\), to
determine the target marginal PMFs \(f_i\). To complete specification of
the simulation algorithm inputs, we estimate the Spearman correlation
matrix \(R_S\) to characterize dependency. With this goal in mind, we
first estimate the desired correlation matrix using the fast
implementation provided by \texttt{bigsimr}:

\begin{Shaded}
\begin{Highlighting}[]
\CommentTok{\# Estimate Spearman\textquotesingle{}s correlation on the count data}
\NormalTok{R\_S }\OtherTok{\textless{}{-}}\NormalTok{ bs}\SpecialCharTok{$}\FunctionTok{cor}\NormalTok{(}\FunctionTok{as.matrix}\NormalTok{(brca1000), bs}\SpecialCharTok{$}\NormalTok{Spearman)}
\end{Highlighting}
\end{Shaded}

Next, we estimate the marginal parameters. We use the method of moments
to estimate the marginal parameters for the multivariate negative
binomial model. The marginal distributions are from the same probability
family (NB), yet they are heterogeneous in terms of the parameters
probability and size \((p_i, n_i)\) for \(i,\ldots,d\). The functions
below support this estimation for later use in \texttt{rvec}.

\begin{Shaded}
\begin{Highlighting}[]
\NormalTok{make\_nbinom\_margins }\OtherTok{\textless{}{-}} \ControlFlowTok{function}\NormalTok{(sizes, probs) \{}
\NormalTok{  margins }\OtherTok{\textless{}{-}} \FunctionTok{lapply}\NormalTok{(}\DecValTok{1}\SpecialCharTok{:}\FunctionTok{length}\NormalTok{(sizes), }\ControlFlowTok{function}\NormalTok{(i) \{}
\NormalTok{    dist}\SpecialCharTok{$}\FunctionTok{NegativeBinomial}\NormalTok{(sizes[i], probs[i])}
\NormalTok{  \})}
  \FunctionTok{do.call}\NormalTok{(c, margins)}
\NormalTok{\}}
\end{Highlighting}
\end{Shaded}

We apply these estimators to the highest-expressing r d genes across the
r nrow(brca1000) patients:

\begin{Shaded}
\begin{Highlighting}[]
\NormalTok{mom\_nbinom }\OtherTok{\textless{}{-}} \ControlFlowTok{function}\NormalTok{(x) \{}
\NormalTok{  m }\OtherTok{\textless{}{-}} \FunctionTok{mean}\NormalTok{(x)}
\NormalTok{  s }\OtherTok{\textless{}{-}} \FunctionTok{sd}\NormalTok{(x)}
  \FunctionTok{c}\NormalTok{(}\AttributeTok{size =}\NormalTok{ m}\SpecialCharTok{\^{}}\DecValTok{2} \SpecialCharTok{/}\NormalTok{ (s}\SpecialCharTok{\^{}}\DecValTok{2} \SpecialCharTok{{-}}\NormalTok{ m), }\AttributeTok{prob =}\NormalTok{ m }\SpecialCharTok{/}\NormalTok{ s}\SpecialCharTok{\^{}}\DecValTok{2}\NormalTok{)}
\NormalTok{\}}
\NormalTok{nbinom\_fit }\OtherTok{\textless{}{-}} \FunctionTok{apply}\NormalTok{(brca1000, }\DecValTok{2}\NormalTok{, mom\_nbinom)}
\NormalTok{sizes }\OtherTok{\textless{}{-}}\NormalTok{ nbinom\_fit[}\StringTok{"size"}\NormalTok{,]}
\NormalTok{probs }\OtherTok{\textless{}{-}}\NormalTok{ nbinom\_fit[}\StringTok{"prob"}\NormalTok{,]}
\NormalTok{nb\_margins }\OtherTok{\textless{}{-}} \FunctionTok{make\_nbinom\_margins}\NormalTok{(sizes, probs)}
\end{Highlighting}
\end{Shaded}

Notably, the estimated marginal NB probabilities \(\{ \hat{p}_i \}\) are
small --- ranging in the interval \([r min(probs) , r max(probs)]\).
This gives rise to highly variable counts and, thus, less restriction on
potential pairwise correlation pairs. Given the marginals, we now
specify targets and check admissibility of the specified correlation
matrix.

\begin{Shaded}
\begin{Highlighting}[]
\CommentTok{\# 1. Mapping step first}
\NormalTok{R\_X }\OtherTok{\textless{}{-}}\NormalTok{ bs}\SpecialCharTok{$}\FunctionTok{cor\_convert}\NormalTok{(R\_S, bs}\SpecialCharTok{$}\NormalTok{Spearman, bs}\SpecialCharTok{$}\NormalTok{Pearson)}
\CommentTok{\# 2a. Check admissibility}
\NormalTok{(is\_valid\_corr }\OtherTok{\textless{}{-}}\NormalTok{ bs}\SpecialCharTok{$}\FunctionTok{iscorrelation}\NormalTok{(R\_X))}
\end{Highlighting}
\end{Shaded}

\begin{verbatim}
## [1] FALSE
\end{verbatim}

\begin{Shaded}
\begin{Highlighting}[]
\CommentTok{\# 2b. compute nearest correlation}
\ControlFlowTok{if}\NormalTok{ (}\SpecialCharTok{!}\NormalTok{is\_valid\_corr) \{}
\NormalTok{  R\_X\_pd }\OtherTok{\textless{}{-}}\NormalTok{ bs}\SpecialCharTok{$}\FunctionTok{cor\_nearPD}\NormalTok{(R\_X)}
  \DocumentationTok{\#\# Quantify the error}
\NormalTok{  targets      }\OtherTok{\textless{}{-}}\NormalTok{ R\_X[}\FunctionTok{lower.tri}\NormalTok{(R\_X, }\AttributeTok{diag =} \ConstantTok{FALSE}\NormalTok{)]}
\NormalTok{  approximates }\OtherTok{\textless{}{-}}\NormalTok{ R\_X\_pd[}\FunctionTok{lower.tri}\NormalTok{(R\_X\_pd, }\AttributeTok{diag =} \ConstantTok{FALSE}\NormalTok{)]}
\NormalTok{  R\_X          }\OtherTok{\textless{}{-}}\NormalTok{ R\_X\_pd}
\NormalTok{\}}
\FunctionTok{summary}\NormalTok{(}\FunctionTok{abs}\NormalTok{(targets }\SpecialCharTok{{-}}\NormalTok{ approximates))}
\end{Highlighting}
\end{Shaded}

\begin{verbatim}
##      Min.   1st Qu.    Median      Mean   3rd Qu.      Max. 
## 0.0000000 0.0001929 0.0004129 0.0005064 0.0007168 0.0069224
\end{verbatim}

As seen above, \(R_X\) is not strictly admissible in our scheme. This
can be seen from the negative result from bs\$iscorrelation(R\_X), which
checks if \(R_X\) is positive-definite with 1's along the diagonal.
However, the approximation is close with a maximum absolute error of r
max( abs(targets - approximates)) and average absolute error of r
mean(abs(targets - approximates)) across the r choose(d,2) correlations.
With the inputs configured, \texttt{rvec} is executed to produce a
synthetic RNA-seq data set:

\begin{Shaded}
\begin{Highlighting}[]
\NormalTok{sim\_nbinom }\OtherTok{\textless{}{-}}\NormalTok{ bs}\SpecialCharTok{$}\FunctionTok{rvec}\NormalTok{(}\DecValTok{10000}\NormalTok{, R\_X, nb\_margins) }
\end{Highlighting}
\end{Shaded}

\begin{figure}
\centering
\includegraphics{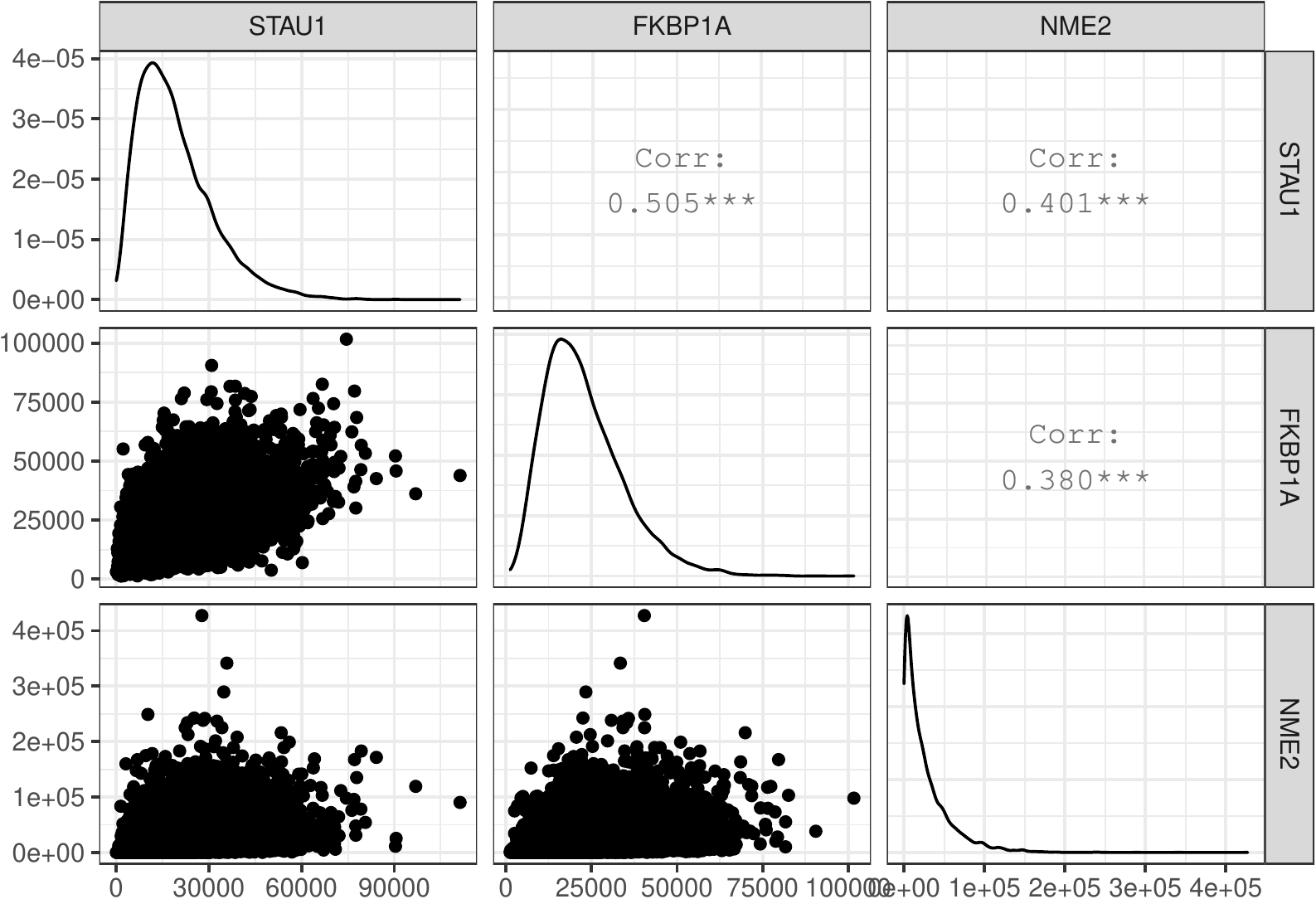}
\caption{\label{fig:ch050-simDataFig}Simulated data for three selected
high-expressing genes generally replicates the estimated data structure.
The data do not exhibit outlying points, but do possess the desired
Spearman correlations, central tendencies, and discrete values.}
\end{figure}

Figure \ref{fig:ch050-simDataFig} displays the simulated counts and
pairwise relationships for our example genes from Table
\ref{tab:ch010-realDataTab}. Simulated counts roughly mimic the observed
data but with a smoother appearance due to the assumed parametric form
and with less extreme points than the observed data in Figure
\ref{fig:ch010-realDataFig}. Figure \ref{fig:ch050-figBRCA} compares the
specified target parameter (horizontal axis) with the corresponding
quantities estimated from the simulated data (vertical axis). The
evaluation shows that the simulated counts approximately match the
target parameters and exhibit the full range of estimated correlations
from the data.

\begin{figure}
\includegraphics[width=0.8\linewidth]{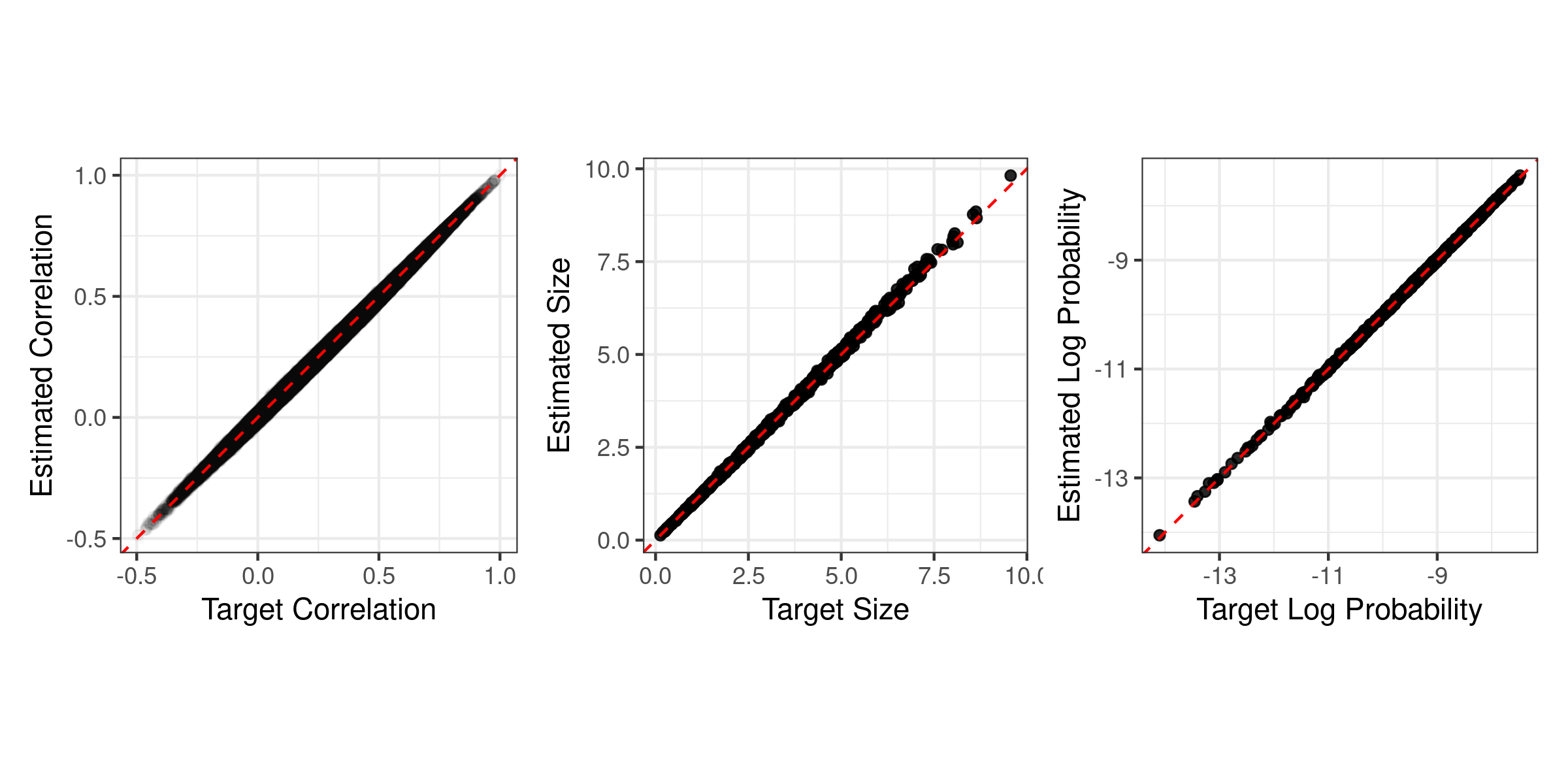} \caption{\label{fig:ch050-figBRCA}Simulated random vectors from a multivariate negative binomial replicate the estimated structure from an RNA-seq data set. The dashed red lines indicate equality between estimated parameters from simulated data (vertical axes) and the specified target parameters (horizontal axes).}\label{fig:ch050-figBRCA}
\end{figure}

\emph{Limitations, conclusions, and recommendations}

The results show overall aquedate simulation performance for our choice
of parameters settings. Our settings were motivated by modeling
high-expressing genes from the TCGA BRCA data set. In general, the
ability to match marginal and dependence parameters depends on the
particular joint probability model. We recommend to evaluate and tune
your simulation until you can be assured of the accuracy.

\hypertarget{discussion}{%
\section{Conclusion and discussion}\label{discussion}}

We developed a general-purpose high-dimensional multivariate simulation
algorithm and provide a user-friendly, high-performance \texttt{R}
package \texttt{bigsimr}, Julia package \texttt{Bigsimr} and Python
package \texttt{bigsimr}. The random vector generation method is
inspired by NORTA (Cario and Nelson 1997) and Gaussian copula-based
approaches Xiao (2017). The major contributions of this work are methods
and software for flexible, scalable simulation of HD multivariate
probability distributions with broad data analytic applications for
modern, big-data statistical computing. For example, one could simulate
high-resolution time series data, such as those consistent with an
auto-regressive moving average model exhibiting a specified Spearman
structure. Our methods could also be used to simulate sparsely
correlated data, as many HD methods assume, via specifying a
\emph{spiked correlation matrix}.

There are limitations to the methodology and implementation. We could
only investigate selected multivariate distributions in our Monte Carlo
studies. There may be instances where the methods do not perform well.
Along those lines, we expect that correlation values close to the
boundary of the feasible region could result in algorithm failure.
Another issue is that for discrete distributions, we use continuous
approximations when the support set is large. This could limit the
scalability/accuracy for particular discrete/mixed multivariate
distributions. Our method would also benefit computationally from faster
Kendall estimation and more finely tuned nearest PD calculation for
non-admissible Kendall matrices.

\hypertarget{refs}{}
\begin{CSLReferences}{1}{0}
\leavevmode\vadjust pre{\hypertarget{ref-BF17}{}}%
Barbiero, Alessandro, and Pier Alda Ferrari. 2017. {``{An R package for
the simulation of correlated discrete variables}.''}
\emph{Communications in Statistics - Simulation and Computation} 46 (7):
5123--40. \url{https://doi.org/10.1080/03610918.2016.1146758}.

\leavevmode\vadjust pre{\hypertarget{ref-Cario1997}{}}%
Cario, Marne C., and Barry L. Nelson. 1997. {``{Modeling and generating
random vectors with arbitrary marginal distributions and correlation
matrix}.''} Technical Report.

\leavevmode\vadjust pre{\hypertarget{ref-Chambers1983}{}}%
Chambers, J. M., W. S. Cleveland, B. Kleiner, and P. A. Tukey. 1983.
\emph{{Graphical Methods for Data Analysis}}. Belmont, CA: Wadsworth
{\&} Brooks.

\leavevmode\vadjust pre{\hypertarget{ref-Chen2001}{}}%
Chen, Huifen. 2001. {``{Initialization for NORTA: Generation of random
vectors with specified marginals and correlations}.''} \emph{INFORMS
Journal on Computing} 13 (4): 312--31.

\leavevmode\vadjust pre{\hypertarget{ref-Chernick2008}{}}%
Chernick, Michael R. 2008. \emph{{Bootstrap Methods: A Guide for
Practitioners and Researchers}}. 2nd ed. Hoboken, NJ.

\leavevmode\vadjust pre{\hypertarget{ref-Conesa2016b}{}}%
Conesa, Ana, Pedro Madrigal, Sonia Tarazona, David Gomez-Cabrero,
Alejandra Cervera, Andrew McPherson, Michal Wojciech Szcześniak, et al.
2016. {``{A survey of best practices for RNA-seq data analysis}.''}
\url{https://doi.org/10.1186/s13059-016-0881-8}.

\leavevmode\vadjust pre{\hypertarget{ref-DH2011}{}}%
Demirtas, Hakan, and Donald Hedeker. 2011. {``{A practical way for
computing approximate lower and upper correlation bounds}.''}
\emph{American Statistician} 65 (2): 104--9.
\url{https://doi.org/10.1198/tast.2011.10090}.

\leavevmode\vadjust pre{\hypertarget{ref-BE07}{}}%
Efron, Bradley. 2007. {``{Correlation and large-scale simultaneous
significance testing}.''} \emph{Journal of the American Statistical
Association} 102 (477): 93--103.
\url{https://doi.org/10.1198/016214506000001211}.

\leavevmode\vadjust pre{\hypertarget{ref-Fasiolo2016}{}}%
Fasiolo, Matteo. 2016. \emph{{An introduction to mvnfast. R package
version 0.1.6.}} \url{https://cran.r-project.org/package=mvnfast}.

\leavevmode\vadjust pre{\hypertarget{ref-GH02}{}}%
Ghosh, Soumyadip, and Shane G Henderson. 2002. {``{Properties of the
Norta method in higher dimensions}.''} \emph{Proceedings of the 2002
Winter Simulation Conference}, 263--69.

\leavevmode\vadjust pre{\hypertarget{ref-higham2002computing}{}}%
Higham, Nicholas J. 2002. {``Computing the Nearest Correlation
Matrix---a Problem from Finance.''} \emph{IMA Journal of Numerical
Analysis} 22 (3): 329--43.

\leavevmode\vadjust pre{\hypertarget{ref-K58}{}}%
Kruskal, William H. 1958. {``{Ordinal measures of association}.''}
\emph{Journal of the American Statistical Association} 53 (284):
814--61. \url{https://doi.org/10.1080/01621459.1958.10501481}.

\leavevmode\vadjust pre{\hypertarget{ref-LH75}{}}%
Li, Shing Ted, and Joseph L. Hammond. 1975. {``{Generation of
pseudorandom numbers with specified univariate distributions and
correlation coefficients}.''} \emph{IEEE Transactions on Systems, Man
and Cybernetics} 5: 557--61.
\url{https://doi.org/10.1109/TSMC.1975.5408380}.

\leavevmode\vadjust pre{\hypertarget{ref-Li2019gpu}{}}%
Li, Xiang, A. Grant Schissler, Rui Wu, Lee Barford, Jr. Harris, Fredrick
C., and Frederick C. Harris. 2019. {``{A Graphical Processing Unit
accelerated NORmal to Anything algorithm for high dimensional
multivariate simulation}.''} \emph{Advances in Intelligent Systems and
Computing}, 339--45. \url{https://doi.org/10.1007/978-3-030-14070-0_46}.

\leavevmode\vadjust pre{\hypertarget{ref-MB13}{}}%
Madsen, L., and D. Birkes. 2013. {``{Simulating dependent discrete
data}.''} \emph{Journal of Statistical Computation and Simulation} 83
(4): 677--91. \url{https://doi.org/10.1080/00949655.2011.632774}.

\leavevmode\vadjust pre{\hypertarget{ref-MK01}{}}%
Mari, Dominique Drouet, and Samuel Kotz. 2001. \emph{{Correlation and
dependence}}. World Scientific.

\leavevmode\vadjust pre{\hypertarget{ref-muino2006gs}{}}%
Muino, JM, Eberhard O Voit, and Albert Sorribas. 2006.
{``GS-Distributions: A New Family of Distributions for Continuous
Unimodal Variables.''} \emph{Computational Statistics and Data Analysis}
50 (10): 2769--98.

\leavevmode\vadjust pre{\hypertarget{ref-Nelsen2007}{}}%
Nelsen, Roger B. 2007. \emph{{An Introduction to Copulas}}. 2nd ed. New
York: Springer Science {\&} Business Media.

\leavevmode\vadjust pre{\hypertarget{ref-Nik13a}{}}%
Nikoloulopoulos, Aristidis K. 2013. {``{Copula-based models for
multivariate discrete response data}.''} In \emph{Lecture Notes in
Statistics: Copulae in Mathematical and Quantitative Finance}, 213th
ed., 231--49. Heidelberg: Springer.

\leavevmode\vadjust pre{\hypertarget{ref-Park1996}{}}%
Park, Chul Gyu, Taesung Park, and Dong Wan Shin. 1996. {``{A simple
method for generating correlated binary variates}.''} \emph{American
Statistician} 50 (4): 306--10.
\url{https://doi.org/10.1080/00031305.1996.10473557}.

\leavevmode\vadjust pre{\hypertarget{ref-qi2006quadratically}{}}%
Qi, Houduo, and Defeng Sun. 2006. {``A Quadratically Convergent Newton
Method for Computing the Nearest Correlation Matrix.''} \emph{SIAM
Journal on Matrix Analysis and Applications} 28 (2): 360--85.

\leavevmode\vadjust pre{\hypertarget{ref-Schissler2019}{}}%
Schissler, Alfred Grant, Dillon Aberasturi, Colleen Kenost, and Yves A.
Lussier. 2019. {``{A Single-Subject Method to Detect Pathways Enriched
With Alternatively Spliced Genes}.''} \emph{Frontiers in Genetics} 10
(414). \url{https://doi.org/10.3389/fgene.2019.00414}.

\leavevmode\vadjust pre{\hypertarget{ref-Schissler2018}{}}%
Schissler, Alfred Grant, Walter W Piegorsch, and Yves A Lussier. 2018.
{``{Testing for differentially expressed genetic pathways with
single-subject N-of-1 data in the presence of inter-gene
correlation}.''} \emph{Statistical Methods in Medical Research} 27 (12):
3797--3813. \url{https://doi.org/10.1177/0962280217712271}.

\leavevmode\vadjust pre{\hypertarget{ref-Ubeda-Flores2017}{}}%
Úbeda-Flores, Manuel, and Juan Fernández-Sánchez. 2017. {``{Sklar's
theorem: The cornerstone of the Theory of Copulas}.''} In \emph{Copulas
and Dependence Models with Applications}.
\url{https://doi.org/10.1007/978-3-319-64221-5_15}.

\leavevmode\vadjust pre{\hypertarget{ref-Wang2009b}{}}%
Wang, Zhong, Mark Gerstein, and Michael Snyder. 2009. {``{RNA-Seq: A
revolutionary tool for transcriptomics}.''}
\url{https://doi.org/10.1038/nrg2484}.

\leavevmode\vadjust pre{\hypertarget{ref-Xia17}{}}%
Xiao, Qing. 2017. {``{Generating correlated random vector involving
discrete variables}.''} \emph{Communications in Statistics - Theory and
Methods} 46 (4): 1594--1605.
\url{https://doi.org/10.1080/03610926.2015.1024860}.

\leavevmode\vadjust pre{\hypertarget{ref-XZ19}{}}%
Xiao, Qing, and Shaowu Zhou. 2019. {``{Matching a correlation
coefficient by a Gaussian copula}.''} \emph{Communications in Statistics
- Theory and Methods} 48 (7): 1728--47.
\url{https://doi.org/10.1080/03610926.2018.1439962}.

\leavevmode\vadjust pre{\hypertarget{ref-Yan2007}{}}%
Yan, Jun. 2007. {``{Enjoy the joy of copulas: with a package copula}.''}
\emph{Journal of Statistical Software} 21 (4): 1--21.
\url{http://www.jstatsoft.org/v21/i04}.

\end{CSLReferences}

\bibliographystyle{unsrt}
\bibliography{references.bib}

\end{document}